\documentclass[preprint,12pt]{elsarticle}

\usepackage{amssymb}

\usepackage{amsmath}
\usepackage{lineno}
\usepackage{url}
\usepackage[ruled,longend]{algorithm2e}
\usepackage{multirow}
\usepackage{textgreek}
\usepackage{amssymb}
\usepackage{wasysym}
\usepackage{caption}
\usepackage{subcaption}
\usepackage{footnotebackref}
\usepackage{rotating}
\usepackage{tablefootnote}
\usepackage{threeparttable}
\usepackage{algorithm2e}
\usepackage{algorithmicx}
\usepackage{csquotes}
\usepackage{algorithmicx}
\usepackage{tikz}
\usepackage{graphicx}
\usepackage{threeparttable}
\usepackage{enumitem}
\usepackage{xcolor}

\usepackage{placeins} 

\usetikzlibrary{positioning, shapes, shapes.geometric, shapes.symbols, shapes.arrows, shapes.multipart, shapes.callouts, shapes.misc}

\journal{Journal of Systems and Software}

\begin{document}

\begin{frontmatter}



\title{Development and Benchmarking of Multilingual Code Clone Detector}


\author[meidai,*]{Wenqing Zhu}
\author[ritsumeikann]{Norihiro Yoshida}
\author[shimane]{Toshihiro Kamiya}
\author[kyotoI]{Eunjong Choi}
\author[meidai]{Hiroaki Takada}

\affiliation[meidai]{organization={Graduate School of Informatics, Nagoya University},
            addressline={Furo-cho, Chikusa-ward}, 
            city={Nagoya-City},
            postcode={464-8601}, 
            state={Aichi},
            country={Japan}}

\affiliation[ritsumeikann]{organization={Graduate School of Information Science and Engineering, Ritsumeikan University},
            addressline={1-1-1 Noji-higashi}, 
            city={Kusatsu},
            postcode={525-8577}, 
            state={Shiga},
            country={Japan}}
\affiliation[shimane]{organization={Interdisciplinary Faculty of Science and Engineering, Shimane University},
addressline={1060 Nishikawatsu-cho}, 
city={Matsue},
postcode={690-8504}, 
state={Shimane},
country={Japan}}
\affiliation[kyotoI]{organization={Faculty of Information and Human Sciences, Kyoto Institute of Technology},
addressline={Matsugasaki, Sakyo-ku}, 
city={Kyoto},
postcode={606-8585}, 
state={Kyoto},
country={Japan}}

\affiliation[*]{Corresponding author: Wenqing Zhu, zhuwqing1995@ertl.jp}{}

\begin{abstract}
The diversity of programming languages is growing, making the language extensibility of code clone detectors crucial. 
However, this is challenging for most existing clone detection detectors because the source code handler needs modifications, 
which requires specialist-level knowledge of the targeted language and is time-consuming.  
Multilingual code clone detectors make it easier to add new language support by providing syntax information of the target language only. 
To address the shortcomings of existing multilingual detectors for language scalability and detection performance, 
we propose a multilingual code block extraction method based on ANTLR parser generation,
and implement a multilingual code clone detector (MSCCD), which supports the most significant number of languages currently available and has the ability to detect Type-3 code clones. 
We follow the methodology of previous studies to evaluate the detection performance of the Java language. 
Compared to ten state-of-the-art detectors, MSCCD performs at an average level while it also supports a significantly larger number of languages.
Furthermore, we propose the first multilingual syntactic code clone evaluation benchmark based on the CodeNet database.
Our results reveal that even when applying the same detection approach, performance can vary markedly depending on the language of the source code under investigation.
Overall, MSCCD is the most balanced one among the evaluated tools when considering detection performance and language extensibility.  

\end{abstract}





\end{frontmatter}


\section{Introduction}
\label{Sec1}

Code clones are pairs of code snippets that are identical or similar. 
They are mainly introduced by copy-and-paste programmers while using code generators and so on can also create them \cite{roy2007survey}. 
Although code cloning can increase the speed of software development, 
code clones are considered bad smells that reduce code quality \cite{Fowler1999}. 
They may lead to the propagation of defects
and difficulty in software maintenance because any modification would have to be adopted to their clones \cite{roy2007survey}. 
Therefore, code clones require to be detected automatically and managed.

Since the 1990s, various code clone detectors that have made good progress in scalability and accuracy (recall and precision) have been proposed.
State-of-the-art tools are known to achieve clone detection for repositories with 100MLOC and provide high-accuracy near-miss clone detection \cite{SAGA,srcClone,ccstokener,SourcererCC}.

Similarly, language extensibility is also a critical evaluation index for clone detection approaches \cite{roy2007survey}. 
The language extensibility of a code clone detector refers to its ability to support multiple languages and extend to new languages.
The proposition of new programming languages and the release of existing languages are frequent. 
According to our previous survey \cite{MSCCD}, 
most popular languages maintain major releases (more likely to include syntax updates) at least once every two years, 
including languages that have been available for less than eight years.
However, it is impractical for existing code clone detectors to support multiple languages because they have to adapt their source-code handler program, 
which is a significant burden to researchers or users requiring specialized knowledge or time.
Hence, most code clone detectors only support a limited range of programming languages and often lack support for the latest language versions.
Among the 13 clone detectors proposed from 2013 to 2018, 
12 mainly supported C or Java \cite{review2019}.

To address the issue of insufficient language extensibility,
several multilingual code clone detectors have been developed \cite{NICAD,CCFinderSW}. 
These detectors allow new language support options by simply modifying the configuration files without altering the tool's code by reusing existing grammar definition files.
NiCad \cite{NICAD}, a well-known clone detector, allows users to specify the analysis method for each language based on TXL\cite{cordy2006txl}.
However, writing TXL syntax for a new language remains challenging for users without expert-level knowledge.
Additionally, creating new TXL files for code clone detection is unrealistic for maintenance developers because of time constraints.
Therefore, the language scalability of NiCad under application is lower than that in theory.
A subsequent of the CCFinder series, known as CCFinderSW \cite{CCFinderSW},  has achieved a lexical analysis mechanism 
by providing the grammar of comments, identifier names, and keywords of the targeted language. 
However, CCFinderSW does not support some languages, such as Lua, and cannot detect a majority of Type-3 clones (details in Section \ref{Sec6}).

This study aims to develop and evaluate a multilingual code clone detection technique.
We address the shortcomings of NiCad and CCFinderSW 
and implement a multilingual syntactic code clone detector (MSCCD), 
which can detect Type-3 clones for the most popular languages.
We answer the following research questions using MSCCD and other code clone detectors.
\begin{itemize}
    \item \textbf{RQ1}: How does MSCCD perform in recall, precision, and scalability compared with the state-of-the-art tools?
    \item \textbf{RQ2}: How do the performances of code clone detection approaches differ for various languages?
    \item \textbf{RQ3}: How extensible are the existing source-code handlers of the multilingual code clone detectors?
\end{itemize}

The language processing mechanism of MSCCD is implemented based on ANTLR, a widely-used LL parser generator\cite{ANTLR}. 
To support a new language, MSCCD only requires modification of the configuration file to import the ANTLR grammar definition for the target language. 
The definitions for most common languages are available in Grammars-v4\footnote{\url{https://github.com/antlr/grammars-v4}}. 
The MSCCD uses the ANTLR-generated parser for each source file to obtain a parse tree (PT), from which code blocks are extracted as clone candidates. 
This study presents a simplified parse tree (SPT) algorithm designed explicitly for MSCCD and a keyword-based filter to remove invalid results from the code-block extraction process. 
Subsequently, MSCCD converts the code blocks into token bags and detects clones by computing overlap similarity between the token bags. 
This approach has shown excellent accuracy and scalability in a previous study \cite{SourcererCC}.

Due to lacking a multilingual clone detection evaluation benchmark, 
most existing studies have mainly focused on Java for evaluation \cite{SourcererCC,CCAligner,srcClone,NIL,SAGA}. 
However, there is no general method for predicting if a technique's performance would be consistent across different languages.
As a solution, we propose two benchmarks to evaluate the precision and recall of clone detectors for four languages: Java, Python, C, and C++.
We used the CodeNet database \cite{CodeNet}, 
which mainly contains data from two online judge systems (OJ system) for competitive programming: AIZU OJ\footnote{\url{https://judge.u-aizu.ac.jp/}} and AtCoder\footnote{\url{https://atcoder.jp/}}. 
Among the data, code pairs accepted by the same OJ problem were regarded as code clones (at least Type-4 clones). 
Based on this, we generated a benchmark by selecting 12 sub-datasets that could reflect those parts of the edited distance similarity covered by the target clone detector. 
Thus, we generated a benchmark to check the correctness of the target clone detector by randomly selecting OJ problem pairs.
When OJ problems differ, if parts of the problems are similar, code clone detectors at the level of code snippets or functions might consider these parts as code clones. 
We selected 90 OJ problem pairs and categorized them into three groups based on their Jaccard text similarity of problem description text: 0-30\%, 30-60\%, 60-90\%, 
and use these as datasets to evaluate the precision of each tool.
In addition to MSCCD, we evaluated CCFinderSW, NiCad, and SourcererCC by these benchmarks. 
The benchmark can be accessed at GitHub \footnote{\url{https://github.com/zhuwq585/MCCD_Benckmarking}}.

For this study, we followed the BigCloneBench-based method and evaluated the recall and precision of the MSCCD for comparison with numerous existing clone detectors \cite{SourcererCC}.
We also compared MSCCD with the other multilingual code clone detectors, NiCad and CCFinderSW, regarding language scalability and detection accuracy. 
We showed that MSCCD was the most balanced detector because it supported the greatest number of languages and had high detection accuracy. 
However, MSCCD is still not the perfect choice. 
First, the performance of the ANTLR-generated parser has to be improved. 
It is prone to many parsing errors, affecting the detection results. 
Second, MSCCD does not have a near-perfect recall in ST3 (defined in Section \ref{Sec2}) and later categories and cannot detect large-variance clones. 
In these writers' view, an intermediate layer is required for clone detection in the future, as it advances the utilization of the official parser by specifying the standard output format. 
Furthermore, the intermediate layer allows researchers to apply their techniques to more programming languages easily. 
The language portability or discrimination problem can be recognized and dealt with for code clone detection techniques, including language-dependent parts.

The main contributions of this study are listed as follows. 
\begin{itemize}
\item We implemented a tool known as MSCCD\footnote{\url{https://zhuwq585.github.io/MSCCD/}} \cite{MSCCD}, which can detect Type-3 clones for a target language by providing its ANTLR grammar-definition file. 

\item Based on the evaluations, MSCCD supported most of the widely used languages and had a competitive performance with the state-of-the-art Type-3 clone detectors.

\item We constructed two clone-detector evaluation benchmarks for Java, C, C++, and Python using the CodeNet dataset. One evaluated the recall and classified the results from Type-1\&2 and Type-3\&4. The other one evaluated the precision of the clone detectors. 

\item A comparative study using the proposed multilingual benchmarks showed that recall and precision of clone detection implementations can also vary with language.

\item MSCCD was the most balanced detector compared to the other multilingual clone detectors. In the future, additional studies are required to advance the use of official parsers and more multilingualization of detection practices.

\end{itemize}

The rest of this paper is organized as follows. 
Section \ref{Sec2} introduces the background and our motivations.
Section \ref{Sec3} discusses the implementation of MSCCD.
Section \ref{Sec4} introduces evaluations based on BigCloneBench.
Section \ref{Sec5} describes the implementation of our multilingual code clone detector evaluation benchmark and the results of evaluating four clone detectors.
Section \ref{Sec6} presents a comparative study of three multilingual code clone detectors in their language extensibilities.
Section \ref{Sec7} discusses the better multilingual code clone detection approach, the real-world applicability of MSCCD, and threats to validity. 
Section \ref{Sec8} introduces related works, while Section \ref{Sec9} presents our conclusions and future plans.

\section{Background}
\label{Sec2}

\subsection{Terminology}

In this paper, we have used the following terminologies and considered their respective definitions. 

\textbf{Token bag}: A multi-set of keywords, identifiers, and literals.

\textbf{Code segment}: A section of contiguous lines of code defined by the quaternion ($f$,$s$,$e$,$g$), with the source file $f$, start line index $s$, stop line index $e$, and granularity value $g$.

\textbf{Code block}: A code segment whose sentences are grouped by one grammar rule.

\textbf{Composition}: The composition of a language is defined as a code block corresponding to a party of grammar rules. Classes, condition statements, loop statements, or functions can form a composition. It is mainly used to evaluate the ability of MSCCD to generate token bags, as discussed in Section \ref{Sec6}.

\textbf{Types of code clones} \cite{roy2007survey,BCB}: 
\begin{itemize}
    \item \textbf{T1 (Type-1)}: These have identical code segments, except for the differences in white space, layout, and comments.
    \item \textbf{T2 (Type-2)}: These have identical code segments, except for the differences in identifier names and literal values, along with the T1 clone differences.
    \item \textbf{T3 (Type-3)}: These have syntactically similar code segments that differ at the statement level. The segments that have statements added, modified, and/or removed with respect to each other and the T1 and T2 clone differences. T3 clones are classified by their syntactic similarity $\theta$:
    \begin{itemize}
        \item \textbf{VST3 (Very strongly Type-3)}: $\theta\in [90,100)$
        \item \textbf{ST3 (Strongly Type-3)}:  $\theta\in [70,90)$
        \item \textbf{MT3 (Moderately Type-3)}:  $\theta\in [50,70)$
        \item \textbf{WT3 (Weakly Type-3)}:  $\theta\in [0,50)$
    \end{itemize}
    \item \textbf{T4 (Type-4)}: These have syntactically dissimilar code segments that implement the same functionality.
\end{itemize}

\subsection{Multilingual Code Clone Detection}

Programming languages may not be able to satisfy all requirements, including writability, extensibility, and cost \cite{ProgrammingLanguageEvolution,LanguageWars}.
Therefore, several languages have been developed according to different purposes.
Besides, the existing languages are updated frequently to keep pace with improvements. 
Our previous survey showed that for the ten most popular languages in September 2021 (ranked by PYPL\footnote{\url{https://pypl.github.io/}}), 
nearly all languages had major releases at a frequency higher than once every two years \cite{MSCCD}, providing a high diversity of programming languages.

\begin{figure*}
    \graphicspath{{./img/}}
        \includegraphics[width=0.97\textwidth]{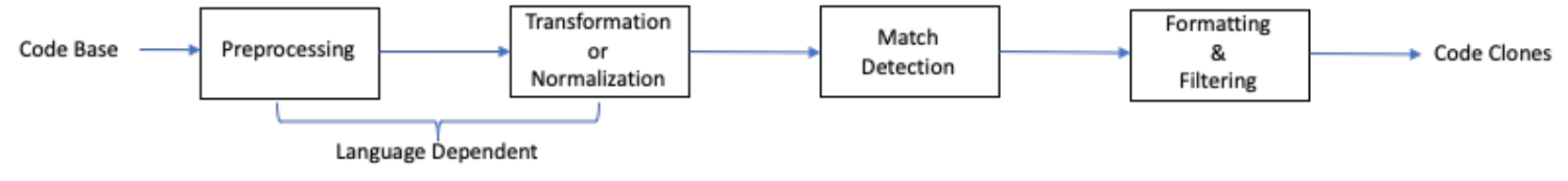}
    \caption{General clone detection process}
    \label{fig:CloneDetectionProcess}
\end{figure*}

Code clone detectors have to deal with a variety of situations.
As shown in Fig. \ref{fig:CloneDetectionProcess}, 
pre-processing, transformation, and normalization of the detection 
process are language-dependent \cite{roy2007survey}. 
When adding a new programming language support to a code clone detector, most operations in these processes require a specific understanding of the language's syntax, such as rules for removing comments, pretty printing source code, tokenization, and parsing.
Therefore, an appropriate source code handler has to be developed for the new language to achieve pre-processing, transformation, and normalization. 
Correspondingly, when the release of an existing language includes syntax modifications, 
the developer must also modify the source code handler to support the new syntaxes. 
Therefore, most code clone detectors cannot easily and quickly add support for a new language or new version of an existing language.

Another important reason is the trade-off between language extensibility and the detection accuracy of clone detectors. 
Current studies are focused on improving the detection accuracy (recall and precision) of T3 and T4 clones \cite{srcClone,wu2020scdetector,ccstokener}.
Other code clone terms, such as gapped clones \cite{CCAligner, NIL} and near-miss clones \cite{saini2018oreo}, are also garnering attention.
Detecting such more difficult-to-detect parts, which require more and higher-dimensional data and more complex pre-processing processes, 
is particularly challenging in multilingual contexts, as demonstrated by code clone detection approaches based on program dependence graph (PDG) or program slicing \cite{srcClone,ccsharp}.
Naturally, researchers are more concerned about the accuracy of their tools than with the number of languages supported by the tool.

Owing to the reasons mentioned above, the language extensibility of existing tools requires improvement.
In the 13 code clone detectors proposed from 2013 to 2018, 12 tools supported a limited range of programming languages, including Java and C/C++ \cite{review2019}.
In addition, the existing evaluation benchmarks for code clone detection using C and Java as the target languages and lack support for other languages \cite{BCB,Svajlenko2021}.
These two factors form a vicious circle that hinders research on clone detection approaches considering linguistic diversity. 
Researchers are highly willing to implement their clone detection approach with benchmark-supported languages in this context. 
Therefore, benchmarks for other languages are unnecessary because most detectors only support a tiny range of benchmark-supported languages.

The lack of support for other languages has caused the research on code clones targeting other languages to stay caught up in progress.
Researchers are yet to confirm whether the published detection methods perform as well for other languages. 
In addition, the focus of detection varies depending on the detection method. 
(For example, some methods are good at detecting gapped clones, while others are good at near-miss clones.) 
With limited choice, or even only their implementation,
studies on other languages often apply only one detection method, 
which may affect their effectiveness \cite{cheung2016jsClone, rustClone_pizzolotto2022}.

Given these challenges, several multilingual clone detectors have been developed. 
A multilingual clone detector is one that provides a multilingual code-handling mechanism that can change language support without modifying the original program, which achieves high language extensibility.
While tools like NiCad\cite{NICAD}, CCFinderSW\cite{CCFinderSW}, and some text-based language-independent approaches \cite{languageIndependentDucasse1999} have met these definitions, 
their each have their unique strengths and weaknesses.
Issues such as T3 clone detection, low precision, and language extensibility (as discussed in Section \ref{Sec7}) vary between tools.
Thus, achieving the best balance between detection capability and language extensibility proves challenging.
In this study, we implement a relatively balanced multilingual code clone detector and evaluate it to derive the direction of the best multilingual code clone detector.

\section{Proposed Tool: MSCCD}
\label{Sec3}
This part aims to develop a tool that detects Type-3 clones 
and can be easily expanded for various programming languages.
We chose the most balanced approach in language extensibility and detection ability.
Moreover, we implemented a detector named MSCCD (Multilingual Syntactic Code Clone Detector):

\subsection{Overview of MSCCD}

\begin{figure}

    \resizebox{\textwidth}{!}{
        \begin{tikzpicture}[node distance=5pt]
        \tikzset{
            input/.style = {
            shape= tape,
            minimum size = 1pt,
            draw = black,
            align= center
            },
            process/.style = {
            rectangle,
            inner sep = 2pt,
            minimum size = 1pt,
            draw = black,
            align= center
            },
            data/.style = {
            trapezium,
            trapezium left angle=70,
            trapezium right angle=-70,
            minimum size = 1pt,
            draw = black,
            align= center
            },
            mrk/.style = {
            align= center
            }
        }
        
        \node[input](grammar) at (-0.3,0.7) {\small Grammar \\ Definition};
        \node[process](ANTLR) at (-0.2,-0.8) {\small ANTLR Parser\\Generation};
        \node[data](parser) at (3.6,-0.8){\small Parser + Code-block\\Extractor};
        \node[input](keywords) at (10,0.8) {\small Keywords List};
        \node[input](Source) at (7.3,0.8){\small Source Files};
        \node[process](normalization) at (7.8,-0.8){\small Token Bag \\ Generation};
        \node[data](tokenBags) at (10.6,-0.8) {\small Token\\ Bags};
        \node[process](CloneDetection) at (13,-0.8){\small Clone\\ Detection};
        \node[data](Clone) at (15.5,-0.8) {\small Clone\\ Pairs};
        
        \draw[->] (grammar) -- (ANTLR);
        \draw[->] (ANTLR) -- (parser);
        \draw[->] (parser) -- (normalization); 
        \draw[->] (keywords) -- (normalization);
        \draw[->] (Source) -- (normalization);
        \draw[->] (normalization) -- (tokenBags);
        \draw[->] (tokenBags) -- (CloneDetection);
        \draw[->] (CloneDetection) -- (Clone);
        
        \draw[dotted] (5.95, 1.9) -- (5.95,-1.4);
        \draw[dotted] (11.85,1.7) -- (11.85,-1.4);
        \node[mrk](m1) at (2.4, 1.7) { Code Transformer Generation };
        \node[mrk](m2) at (9, 1.7) { Token Bag Generation };
        \node[mrk](m3) at (14, 1.7) {Clone Detection};
        \node[mrk](m3) at (3.6, 0.05) {Code Transformer};

    \end{tikzpicture}
    }

        \caption{Overview of MSCCD}
    \label{fig:stmProcess}
\end{figure}
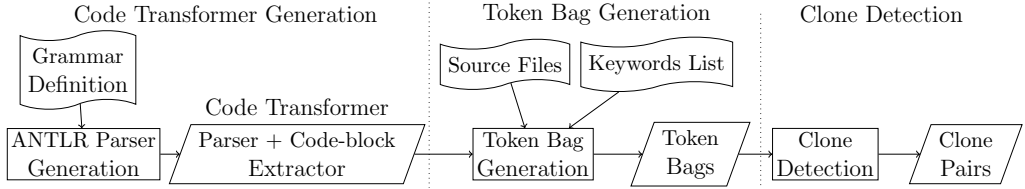

Fig. \ref{fig:stmProcess} presents an overview of MSCCD. 
The system operates three phases: code transformer generation,
token bag generation, and clone detection.
Initially, MSCCD generates a source code transformer for the targeted language using the ANTLR parser generator. 
In the token bag generation phase, the code transformer extracts code blocks and converts the code blocks into token bags. 
Then in the clone detection phase, code clones are identified as similar token bags.

The source code transformer generated in the first phase includes a parser for the targeted language
and a code block extractor. 
ANTLR parser generator allows us to add support for new languages by just providing the grammar definition file of the language without modifying the code of MSCCD. 
Similarly, providing the new version of the syntax description file is sufficient for syntax updates of the existing languages. 
The source code transformation approach used in this study relies on a ``generic tool-generated syntax tree.’’ 
It can be realized through the utilization of any parser generator.
We chose ANTLR4\footnote{\url{https://www.antlr.org}}, an LL-based parser generator, for MSCCD's implementation because of its expansive library of over 150 grammar definition files available in the ``grammars-v4’’\footnote{\url{https://github.com/antlr/grammars-v4}} repository.
Users can directly leverage these grammar definition files without creating new ones.
Furthermore, ANTLR4's rich API helps mitigate implementation challenges.

\subsection{Code Block Extraction using Parse Tree}
\label{Subsec3.2}
\begin{figure}
    \graphicspath{{./img/}}
        \centering
        \includegraphics[width=0.9\textwidth]{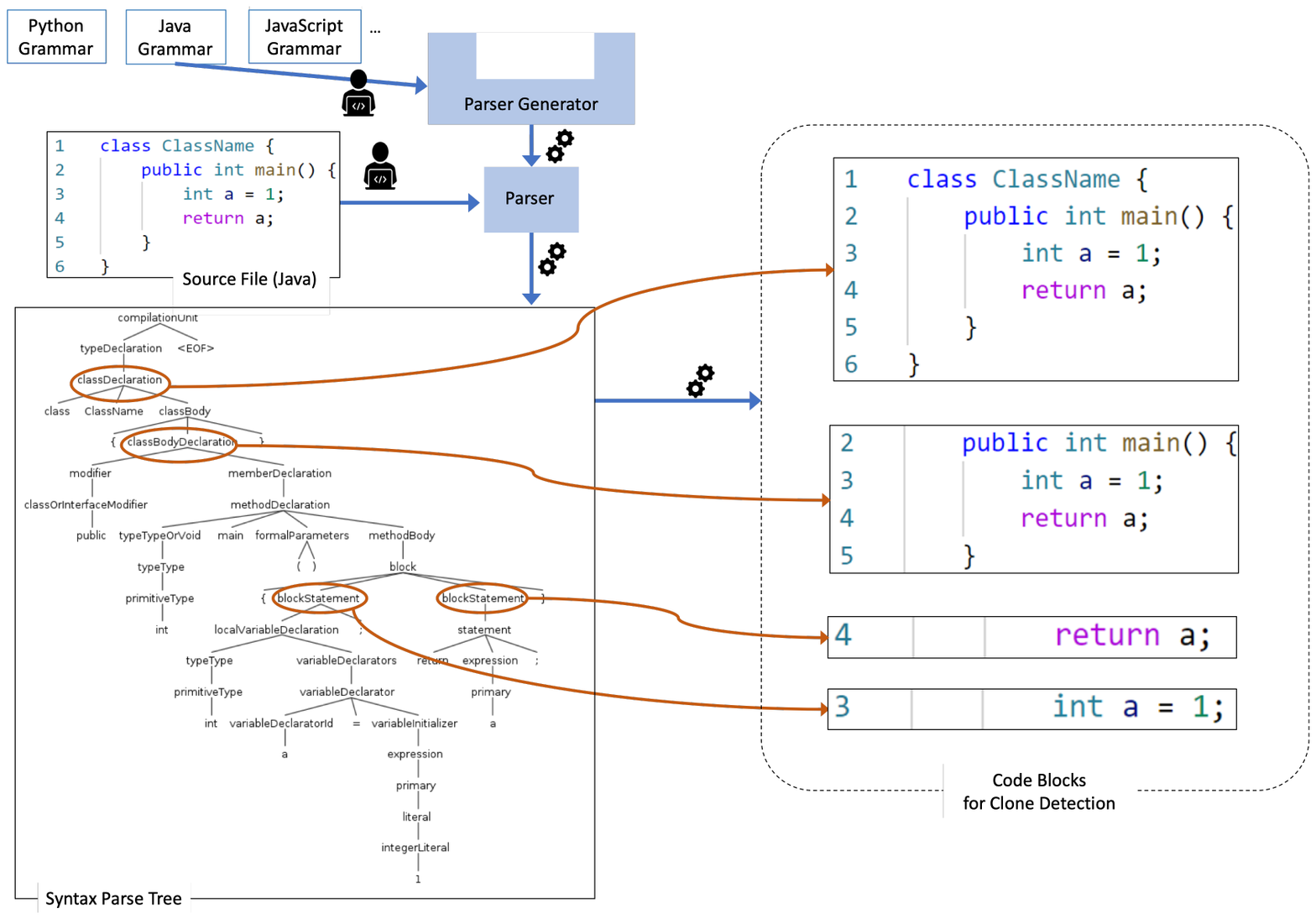}
    \caption{Multilingual Code Block Partition by Parse Tree \cite{MSCCD}}
    \label{fig:BlockPartitionIdea}
\end{figure}

The token bag-based clone detection method has been proven to have excellent detection performance by previous studies\cite{MSCCD,nishi2018scalable}. 
The method can effectively detect Type-3 code clones with high precision and achieve high scalability for repositories with 100MLOC. 
Other candidate methods perform worse or cannot be applied to general parse trees (PTs). 
(Note that in this context, we only generate the bare parse tree and do not perform any language-dependent operations based on the semantics of the target language.)
In the last two phases, 
MSCCD converts each source code file into token bags 
and then detects code clones by similarity comparison.
The details are described in the following two subsections.

\begin{figure}
    \centering
    \begin{tikzpicture}
      \tikzset{
        data/.style = {
        shape= tape,
        minimum size = 15pt,
        draw = black,
        align= center
      },
      treenode/.style = {
        shape=circle,
        inner sep = 2pt,
        minimum size = 2pt,
        draw = black,
        align= center
      },
      noneNode/.style = {
        draw = white,
        shape=circle,
        inner sep = 2pt,
        minimum size = 2pt,
        align= center
      },
      mrk/.style = {
        align= center
      }
      }
      \footnotesize
  
      \node[noneNode](a00) at (0, 0) {};
  
      \node[treenode](a00) at (1.8, 1.5) {0};
      \node[treenode](a10) at (1.8, 0.9) {1};
      \node[treenode](a20) at (1.2, 0.3) {2};
      \node[treenode](a21) at (2.4, 0.3) {3};
      \node[treenode](a30) at (0.8, -0.3) {4};
      \node[treenode](a31) at (1.4, -0.3) {5};
      \node[treenode](a32) at (1.9, -0.3) {6};
      \node[treenode](a33) at (2.4, -0.3) {7};
      \node[treenode](a34) at (2.9, -0.3) {8};
      \node[treenode](a40) at (2.1, -0.9) {9};
      \node[treenode](a41) at (2.7, -0.9) {10};
      
      \draw[-] (a00) -- (a10);
      \draw[-] (a10) -- (a20);
      \draw[-] (a10) -- (a21);
      \draw[-] (a20) -- (a30);
      \draw[-] (a20) -- (a31);
      \draw[-] (a21) -- (a32);
      \draw[-] (a21) -- (a33);
      \draw[-] (a21) -- (a34);
      \draw[-] (a33) -- (a40);
      \draw[-] (a33) -- (a41);
      
      \node[treenode](b00) at (5.0, 1.5) {0};
      \node[treenode](b10) at (4.4, 0.5) {2};
      \node[treenode](b11) at (5.6, 0.5) {3};
      \node[treenode](b20) at (5.6, -0.5) {7};
  
      \draw[-] (b00) -- (b10);
      \draw[-] (b00) -- (b11);
      \draw[-] (b11) -- (b20);
  
      \node[noneNode](s0) at (7, 1.5) {0};
      \node[noneNode](s1) at (7, 0.5) {1};
      \node[noneNode](s2) at (7, -0.5) {2};
  
      \draw[dotted] (5.4,1.5 ) -- (6.8,1.5 );
      \draw[dotted] (6,0.5 ) -- (6.8,0.5 );
      \draw[dotted] (6,-0.5 ) -- (6.8, -0.5 );
  
      \node[mrk](m1) at (1.8, 2) {PT};
      \node[mrk](m2) at (5.0, 2) {SPT};
      \node[mrk](m3) at (7, 2) {granularity value};
  
      \draw[->] (m1) -- (m2);
    \end{tikzpicture}
    \begin{tablenotes}
        \centering

      \item *: The minimum token is set to 2.
    \end{tablenotes}
    \caption{ Simplification of a Parse Tree }
    \label{fig:blockExtraction}

  \end{figure}
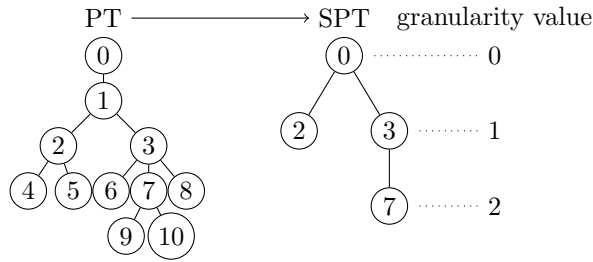

Extracting code blocks from source code needs careful consideration in multilingual code clone detection.
Existing code clone detectors often identify blocks with language-dependent semantics.
However, this method could be better in our case as it requires substantial adjustments for each language.
To circumvent this issue, we propose representing a semantic code block as each subtree in a PT.
Fig. \ref{fig:BlockPartitionIdea} illustrates the process of code block partition.
A PT is an ordered tree representing the process of grammar derivation.
Each non-leaf node in the PT corresponds to a grammar rule, while each leaf node corresponds to a terminator (lexical units) and a word in the source code.
Each subtree whose vertex is a non-leaf node in the PT represents a contiguous part of the source code and corresponds to a rule in the grammar definition. 
These contiguous parts of code can be functions, code blocks, and statements that are the object of clone detection comparisons.
Code blocks can be extracted from the language-independent PT by referring to these subtrees.

\FloatBarrier

However, extracting code blocks corresponding to every subtree for clone detection is insufficient.
Plural subtrees may correspond to identical code blocks in PT, shown as Nodes 0 and 1 in the PT of Fig. \ref{fig:blockExtraction}. 
In the syntax tree of Figure \ref{fig:BlockPartitionIdea}, there is a real example: the node $typeDeclaration$ near the root and its child node $classDeclaration$ correspond to exactly identical leaf nodes.
Grammar rules such as ``$statements: statement+$’’ are common. 
If the non-terminal $statement$ on the right side matches only once in the derivation, 
the final PT will correspond to the same terminators (the same code block) for subtrees with $statements$ and $statements$ as the vertex. 
Besides, if a part of the subtrees corresponds to smaller code blocks or even a part of a statement  
(e.g., Node 9 and 10 in Fig. \ref{fig:blockExtraction}), 
this part should also be excluded from the comparison of clone detection.

\begin{algorithm}
    \caption{Parse Tree Simplification}
    \label{alg:simplify}
      \LinesNumbered
    \SetKw{merge}{merge}
    \SetKw{delete}{delete}
    \KwIn{ $T$ is a PT, each tree node contains an attribute $size$ representing its token number and an attribute $child$ containing child nodes; $mSize$ is the configured minimum size }
    \KwOut{ An SPT }
  
    \SetKwProg{Fn}{Function}{:}{end}
    \SetKwComment{Comment}{//}{}
  
    \Fn{ParseTreeSimplification{\upshape(}$T$, $mSize${\upshape)}}{ 
      \ForEach{{\upshape tree node} $n$ {\upshape in pre-order traversal of} $T$ }{
        \While{$n${\upshape .length} == $n${\upshape .child[0].length}}{
          \merge($n$, $n$.child[0])
        }
       
      \If{$n${\upshape .size} \textless\  $mSize$}{
        \ForEach{{\upshape child node} $cn$ {\upshape from} $n$}{
          \delete($cn$)
        }
      }
      }
      \Return{$T$}
    }
  
\end{algorithm}

To eliminate these invalid code blocks, 
we propose algorithm \ref{alg:simplify} to simplify PT and extract code blocks from them.
We traverse each node $n$ the PT in pre-order.
If the number of corresponding leaf nodes of $n$ equals $n$'s first child node, 
this means that the node $n$ only contains one child node $n.child[0]$, and they correspond to the same leaf nodes.
Then, $n$ and $n.child[0]$ are merged into one node (lines 3--5). 
Ideally, $mSize$ represents the minimum number of tokens for a single statement in the program such that all blocks containing more than one complete and consecutive statement are obtained. 
In practical usage, $mSize$ represents the minimum token for clone detection, 
where all code blocks containing tokens less than $mSize$ will not be extracted.
Therefore, if leaf nodes corresponding to $n$ are less than $mSize$, all the child nodes of $n$ are deleted and will not be traversed (lines 6--10).
We call the tree left behind a Simplified Parse Tree (SPT).
The subtrees of all nodes in the SPT corresponding to the source PT are the target of code block extraction.

In Fig. \ref{fig:blockExtraction}, we illustrate an example of SPT and introduce our definition of granularity value: the depth of each SPT node. 
Here, a larger granularity value represents a finer granularity.
Furthermore, nodes having the same granularity value are more likely to represent the precise composition of the language \cite{MSCCD}. 
For example, in Java, nodes having a depth value of one represent classes.

Additionally, for most languages, certain keywords are essential in defining semantic code blocks, such as ``class" for classes, ``def" or ``function" for functions, and ``for" or ``while" for loops. 
These elements are also clone detection targets. 
Nodes in the PT corresponding to these blocks contain a child node that only contains keywords \cite{MSCCD}. 
Taking this into account, we offer a keyword filter to reduce the number of candidates for detecting clones between specific structures like functions.

In the implementation,
the input project $P$ consists of a number of source files $f$, 
that is, $F$: $P = \left \{ f_{0}, f_{1} , ... , f_{n}\right \}$.
MSCCD partitions each source file $f$ into several token bags $Bf_{GV}$ in each granularity (from 0 to GVmax), 
$f = \left \{ Bf_{0},Bf_{1}, ... , Bf_{GVmax}0  \right \}$, where $Bf_{GV} = \left \{ Bf_{GV}0,Bf_{GV}1,...\right \}$.
We also consider overlaps between token bags corresponding to nodes of different granularities as viable candidates for detection.
Algorithm \ref{Alg:TokenBagGeneration} shows the operation on each file $f$.
For each source file $f$, SPT generated by Algorithm \ref{alg:simplify} is given as input.
For each node, $n$ of the SPT, token bags are generated directly from the subtree corresponding to $n$. 
When the keyword filter is activated (a non-empty keywords list is provided), 
the token types of all the leaf nodes contained in the child nodes of the node $n$ are checked. 
If any child nodes of $n$ do not contain only keywords, node $n$ is excluded from the code block extraction result.

\begin{algorithm}
    \caption{Token Bag Generation}
    \label{Alg:TokenBagGeneration}
    \LinesNumbered
    \KwIn{ $T$ is a SPT generated by Algorithm 1. $K$ is the keywords list.  }
    \KwOut{A Collection of token bags}
    
    \SetKwProg{Fn}{Function}{:}{end}
    \SetKwRepeat{Do}{do}{while}
    \SetKwComment{Comment}{//}{}
    
    \Fn{TOKENBAGGENERATION($T$, $K$)}{
      $TargetNodes$,$TokenBags$ = [];
      
      
      \eIf{$K$ == null}{
            \ForEach{TreeNode $n$ in $T$}{
                $TargetNodes$.append($n$);
            }
        }{
            \ForEach{TreeNode $n$ in $T$}{
                \If{KeywordsFilter($n$, $K$) == True}{
                    $TargetNodes$.append($n$);
                }
            }
        }
        
        \ForEach{TreeNode $tn$ in $TargetNodes$}{
            $TokenBags$.append(token bag created from $tn$);
        }
        \Return{$TokenBags$}
    }
    
    \Fn{KeywordsFilter($n$, $K$)}{
          
      \ForEach{TreeNode $cn$ in $n$.childs}{
            $flag = 0$;
            
            \ForEach{LeafNode $t$ corresponds to $cn$}{
                \eIf{$t \in K$}{
                    $flag \mathrel{|}= 1$;
                }{
                    $flag \mathrel{|}= 3$;
                }
            }
            \If{$flag$ == 1}{
                \Return{True};
            }
        }
        \Return{False};
    }
\end{algorithm}

\subsection{Clone Detection}

During clone detection, 
MSCCD calculates the similarity between all candidate token bag pairs.
To calculate the similarity, we adopted the overlap similarity method, which has reported a good recall and scalability by SourcererCC\cite{SourcererCC}.
For a given pair of token bags, overlap similarity is defined as the number of tokens they share divided by the total number of tokens in the larger bag. 
As presented in equation \ref{equ:overlapSimi}, if the similarity between token bags $B_{x}$ and $B_{y}$ is larger than the threshold $\theta$, the bags are judged as a code clone.

\begin{equation}
    \label{equ:overlapSimi}
    OverlapSimilarity \left( B_{x}, B_{y}  \right) = \frac{\left  |B_{x}\cap B_{y}  \right |}{MAX\left ( \left|B_{x} \right|, \left| B_{y}\right| \right )} \geq \theta \\.
\end{equation}

A challenge in such detection lies in identifying suitable candidates for comparison. In general, clone detectors have to compare every possible code segment against every other potential segment \cite{roy2007survey}.
However, as introduced in the Subsection \ref{Subsec3.2}, there may be an overlap between token bags of different granularities. 
Scoping with all token bags can produce multiple meaningless results, 
such as identifying a code block as a clone with its subset. 
The number of token bags generated at multiple granularities is already significantly higher than those generated at a single granularity. 
Moreover, filtering the meaningless results after detection would add much overhead. 
These can lead to scalability problems.

Thus, we choose to group token bags by granularity values and detect code clones within each group,
as similar components are more likely to appear at the same granularity values.
In the case of copy-and-paste, there is a high probability that the newly created clones will have the same granularity values as the source code.
When setting the biggest granularity value as $g_{max}$ and the number of token bags in granularity value $i$ as $N_{i}$, 
the time complexity of candidate comparison is reduced from $O\left ( \left ( \sum_{i=0}^{g_{max}} N_{i} \right  ) ^{2} \right) $ to $O\left ( \sum_{i=0}^{g_{max}} N_{i}^{2} \right )$.
Besides, the detection can be easily parallelized.

Consider the case of code clones at different granularity levels.
Suppose there is a code block $B1$ with granularity value $x$, and another code block $B2$ at granularity value $x+1$.
In this case, it is highly probable that $B2$'s parent block $B2'$ (of granularity level x) is a clone of $B1$, 
unless the similarity threshold is set to 100\%.
Therefore, avoiding comparisons between code blocks of differing granularity values does not significantly decrease recall.

When token bags $B1$ and $B2$, at the same granularity level $x$, are identified as a code clone, 
it is highly likely that their sub-bags at the next granularity level $(x+1)$, namely $B1'$ and $B2'$, will also be code clones.
For example, corresponding parts of a T1 clone must be a T1 clone.
In such a case, MSCCD only reports clones with the smallest granularity value (the coarser granularity).

\section{Evaluation using BigCloneBench}
\label{Sec4}

As mentioned in Section\ref{Sec2}, 
in recent years, most syntactic clone detectors have been evaluated based on BigCloneBench.
To compare with these existing tools, 
we evaluated MSCCD using the same method with the SourcererCC research based on BigCloneBench (IJaDataset) \cite{BCB,IJaDataset,SourcererCC}, which was also used in many studies.
We cited these studies which relate to syntactic code clone detections to compare MSCCD with them\cite{SourcererCC,CCAligner,srcClone,SAGA,NIL}.
Especially, SourcererCC is the best baseline detector when evaluating MSCCD.
Because SourcererCC is well-evaluated and MSCCD uses the same similarity measure as SourcererCC's to verify code clones.

\newcommand{\specialcell}[2][l]{%
  \begin{tabular}[#1]{@{}l@{}}#2\end{tabular}}

\begin{table}
    \centering
    \caption{Types and  Configurations for Target Detectors}
    \label{Table:ToolConfigurationBCB}
    \resizebox{\textwidth}{!}{
        \begin{tabular}{r|l|l}    
            \hline
            Tool & Code Representation  & Configuration  \\ \hline
            MSCCD & token bag  &min tokens: 20, similarity threshold: 70\%, granularity value: 0-4 \\ \hline
            SourcererCC & token bag & min tokens: 50, similarity threshold: 70\%    \\ \hline
            CCFinderX &token sequence &  \specialcell{min tokens: 50,  min token types: 12} \\ \hline
            NIL & token sequence  &\specialcell{N value: 5, verification threshold: 0.7\\filtration threshold: 0.1}\\ \hline
            CCAligner & token sequence   &\specialcell{min lines: 6, similarity threshold: 60\%,  window size: 6, edit distance: 1} \\ \hline
            SAGA & suffix tree of token sequence   &similarity threshold: 70\%, min tokens: 40, function \\ \hline
            iClones & suffix tree of token sequence  &\specialcell{min tokens: 50,   minblock: 20 tokens} \\ \hline
            NiCad & normalized code text &  \specialcell{min lines: 6, similarity threshold: 70\%\\blind identifier normalization, identifier abstraction} \\ \hline
            srcClone & program slice  &similarity threshold: 75\% \\ \hline
            Deckard & AST &  \specialcell{min tokens: 50, similarity threshold: 85\%,  2 token stride} \\\hline
        \end{tabular}
    }
\end{table}

\begin{table}
    \centering
    \caption{Recall Measurements by BigCloneEval}
    \label{Table:RecallInBCB}
    \resizebox{\textwidth}{!}{
        \begin{tabular}{r|ccccc|ccccc|ccccc|c}
            \hline
        \multirow{2}{*}{Tool} & \multicolumn{5}{c|}{All Clones} & \multicolumn{5}{c|}{Intra-Project Clones}  & \multicolumn{5}{c|}{Inter-Project Clones} & \multirow{2}{*}{\specialcell{Citations}} \\ \cline{2-16}
                                       &  T1 &  T2  &  VST3 &  ST3 & MT3 &  T1 &  T2 &  VST3 &  ST3  & MT3 &  T1 & T2 & VST3 &  ST3  & MT3 \\ \hline 
        MSCCD\cite{MSCCD}              & 100 &  98  &  93   &  62  & 6   & 100 & 100 & 100   &  88   & 26  & 100 & 97 & 85   &  48   & 6  & - \\ \cline{17-17}
        SourcererCC\cite{SourcererCC}  & 100 &  98  &  93   &  61  & 5   & 100 & 99  & 99    &  86   & 14  & 100 & 97 & 86   &  48   & 5  & \multirow{5}{*}{\cite{SourcererCC}} \\ 
        CCFinderX\cite{CCFinderX}      & 100 &  93  &  62   &  15  & 1   & 100 & 89  & 70    &  10   & 4   & 98  & 94 & 53   &  1    & 1   \\
        Deckard\cite{jiang2007deckard} & 60  &  58  &  62   &  31  & 12  & 59  & 60  & 76    &  31   & 12  & 64  & 58 & 46   &  30   & 12  \\
        iClones\cite{iClones}          & 100 &  82  &  82   &  24  & 0   & 100 & 57  & 84    &  33   & 2   & 100 & 86 & 78   &  20   & 0   \\
        NiCad\cite{NICAD}              & 100 &  100 &  100  &  95  & 1   & 100 & 100 & 100   &  99   & 6   & 100 & 100& 100  &  93   & 1   \\ \cline{17-17}
        CCAligner\cite{CCAligner}      & 100 &  99  &  97   &  70  & 10  & -   & -   & -     &  -    & -   & -   & -  & -    &  -    & - & \cite{CCAligner}  \\ \cline{17-17}
        srcClone\cite{srcClone}        & 100 &  100 &  99   &  95  & 77  & -   & -   & -     &  -    & -   & -   & -  & -    &  -    & - & \cite{srcClone}   \\ \cline{17-17}
        SAGA\cite{SAGA}          & 100 &  98  &  91   &  45  & 5   & -   & -   & -     &  -    & -   & -   & -  & -    &  -    & - & \cite{SAGA} \\ \cline{17-17}
        NIL\cite{NIL}                  & 100 &  97  &  94   &  67  & 11  & -   & -   & -     &  -    & -   & -   & -  & -    &  -    & - & \cite{NIL} \\ \hline
        \end{tabular}
    }
\end{table}

\subsection{Recall}
We measured the recall of MSCCD using BigCloneEval \cite{BCE}, which implements a framework of BigCloneBench \cite{BCB}. 
BigCloneBench is a benchmark built by real clones collected from IJaDataset\cite{IJaDataset}, 
a large inter-project repository consisting of more than 25,000 open-source Java projects.
BigCloneBench can report the recall of a code clone detector from Type-1 to Type-4. 
Specially, Type-3 clones are divided into VST3, 
ST3, MT3, and WT3 by their syntactical similarity.
The evaluated tools targeted syntactic code clones; hence, WT3 and T4 results were excluded from this experiment.

Table \ref{Table:RecallInBCB} lists the results of recall measurements.
For all clones, MSCCD had a relatively good recall in T1 (100\%), T2 (98\%), and VST3 (93\%) clones. 
The recall dropped for ST3 clones (from 93\% for VST3 to 62\%).
For MT3, WT3, and T4 clones, MSCCD faced detection challenges.
Similar to most tools, MSCCD performed better in detecting intra-project clones. 
In this case, MSCCD had a perfect recall for T1, T2, and VST3 clones (100\%). 
The recall started dropping for the MT3 clones (from 88\% for ST3 to 26\%).
For inter-project clones, the recall of MSCCD was slightly lower than that in the inter-project clones category, while it fell below 50\% for ST3 clones (48\%).

The recall of MSCCD was similar to that of SourcererCC  
because MSCCD used the same similarity with SourcererCC \cite{SourcererCC}.
However, the recall of MSCCD for ST3 and MT3 was slightly higher than that of SourcererCC because MSCCD detected code clones in various granularities. 
For example, in function-level comparisons, a code clone between functions was flagged as a none-clone by SourcererCC and MSCCD. 
However, MSCCD further compared the similarity between the two functional bodies and detected it as a clone at that granularity. 
When the line coverage between the two functional bodies was higher than 70\% for a metafunction, 
BigCloneEval still considered the clone to be successfully detected.

Most tools perform well in detecting T1, T2, and VST3 code clones, with seven of 10 tools having a recall of at least 91\%. A significant difference was reflected in detecting ST3 clones. 
For ST3 clones, specialized tools (srcClone and NiCad) achieved a recall rate of 95\%, 
while CCFinderX, a tool famous for accurately detecting T2 clones, achieved only approximately 15\%.
MSCCD had an average level of 62\% recall for ST3 clones, which ranked fifth. 
Only SrcClone was able to efficiently detect MT3 clones, achieving a recall rate of 77\%.
The recall for MT3 clones by other tools was no more than 12\%.

\label{Sec4.1}

\subsection{Precision}
\label{Sec4.2}
Automatic precision measurement is still an open question because of the need for high-quality benchmarks. 
A general method randomly selects a sample of detected results and involves manual checking for the correctness of each reported clone \cite{SourcererCC}.
Following this approach, we randomly selected 400 clone pairs from the recall measure experiment discussed in Subsection \ref{Sec4.1}, 
and equally distributed them among five judges to check the correctness \cite{MSCCD}.
For the reported pairs that were hard to judge, the pairs were marked as unknown and counted as false positives.
Owing to the randomness of result sampling, repetition of the experiment would lead to different results. 
 Therefore, we cited multiple results of the target detectors evaluated by the same experiment with the same detector configuration. 
The true performance of the target detector was somewhere in between.

The biggest challenge with this experiment was the need for a larger sample size.
In the case of MSCCD, 61917 clones included in BigCloneBench were detected. 
However, MSCCD reported results that were multiple times the number of clones in different granularities, which were not collected in BigCloneBench.
The smaller the sample size, the greater the likelihood of being affected by randomness. 
Therefore, if the random sampling precision experiment was conducted multiple times, 
results with higher means and fewer fluctuations over multiple trials represented higher precision. 
Furthermore, we conducted the same experiment in our original paper \cite{MSCCD} again for MSCCD.
The results are listed in Table \ref{Table:PrecisionInBCB}.
We observed that the result for MSCCD was 93\%, approximately the same as the average of multiple results of SourcererCC (91\%). 
Among other tools with multiple results, 
NiCad and CCAligner were considered less accurate than SourcererCC and MSCCD because of their lower average values, 
although they also included a high result of more than 90\%.

Another factor that could affect the validity of the comparison was the difference in detection range. 
For example, MSCCD reported clones at multiple granularities such as file, block, and function. 
False positives were more likely to occur between larger code segments. 
The probability of smaller clones, such as loops in the sample, was better, 
which could make the evaluation results higher than the actual value.
Therefore, some studies set the object range above 10 LOC \cite{CCAligner,SourcererCC}.
In this group, MSCCD also ranked second among the seven tools.

\begin{table}
    \centering
    \caption{Precision Measured by Random Sampling Test for BigCloneBench}
    \label{Table:PrecisionInBCB}

    \begin{tabular}{c|c|c|c}
        \hline
        Tool        & \specialcell{Precision \\ \%} & \specialcell{Precision (10 LOC) \\ \%} & \specialcell{Citations} \\ \hline
        MSCCD       & 92-94     & 90  & \cite{MSCCD} \\
        SourcererCC & 83-99     & 86  & \cite{SourcererCC,NIL} \\
        CCFinderX   & 72        & 79  & \cite{SourcererCC} \\
        Deckard     & 28-34     & 30  & \cite{SourcererCC,saini2018oreo} \\
        iClones     & 91        & 93  & \cite{SourcererCC} \\
        NiCad       & 56-99     & 80  & \cite{SourcererCC,saini2018oreo,NIL} \\
        CCAligner   & 34-92     & 83  & \cite{CCAligner,NIL} \\ 
        srcClone    & 90        & -   & \cite{srcClone} \\
        SAGA        & 99        & -   & \cite{SAGA} \\
        NIL         & 94        & -   & \cite{NIL} \\ \hline

    \end{tabular}
\end{table}

\subsection{Execution Time}
\label{Sec4.3}

To evaluate the scalability of MSCCD,
we generated test data from 1KLOC to 100MLOC by randomly selecting files from IJaDataset \cite{IJaDataset}.
We used the ``cloc’’ tool to count the lines of code while excluding blank and comment lines \cite{cloc}.
The experiment was executed on a machine with an i7-8665U CPU and 12GB RAM. The setup was similar to that used in the experiment by Sajnani et al \cite{SourcererCC}.
We counted the execution time of token bag generation and clone detection for each group.
The time consumed for tokenizer generation was not counted because only a single execution was conducted, which only took less than 5 sec.
The configuration of MSCCD was the same as that discussed in Section \ref{Sec4.1}.

\begin{table}
    \caption{Execution Time Measurement}
    \label{Table:scalability}

    \fontsize{8pt}{12pt}\selectfont

    \begin{subtable}[t]{\textwidth}
    \centering
    \caption{Used Machine in Each Detectors' Experiment}
    \label{Table:ParserErrorDetails.a}
           \begin{tabular}{c|c|c}
                \hline
               Detector & Machine & Citations             \\ \hline
                MSCCD & i7-8665U (quad-core) CPU, 12GB RAM & - \\ \hline    
                NiCad & \multirow{5}{*}{3.5GHz quad-core i7 CPU, 12GB RAM} & \multirow{5}{*}{\cite{SourcererCC}} \\ \cline{1-1}
                SourcererCC & &  \\ \cline{1-1}
                CCFinderX  &  &     \\ \cline{1-1}
                Deckard &  &   \\ \cline{1-1}
                iClones &  &   \\ \cline{1-3}
                CCAligner & quad-core CPU, 12GB RAM & \cite{CCAligner} \\ \cline{1-3}
            \end{tabular}
    \end{subtable}

    \hfill
    
    \begin{subtable}[t]{\textwidth}
    \centering
    \caption{Execution Time}
        \resizebox{\textwidth}{!}{
            \begin{tabular}{c|llllll}
            \hline
           Detector & 1K    & 10K    & 100K    & 1M           & 10M                & 100M              \\ \hline
            MSCCD & 5 s & 5 s  & 7 s  & 32 s & 15 m 9 s &  MEMORY  \\ \hline
            NiCad & 1 s & 4 s & 21 s & 4 m 1 s & 11 h 42 m 47 s & INTERNAL LIMIT   \\ \hline
            SourcererCC & 3 s & 6 s  & 15 s  & 1 m 30 s & 32 m 11 s & 1 d 12 h 54 m     \\ \hline
            CCFinderX & 3 s & 4 s  & 21 s  & 2 m 18 s & 28 m 51 s & 3 d 5 h 49 m         \\ \hline
            Deckard & 2 s & 9 s  & 1 m 34 s  & 1 h 12 m 3 s & MEMORY & -         \\ \hline
            iClones & 1 s & 1 s  & 2 s  & MEMORY & - & -         \\ \hline
            CCAligner & 1 s & 1 s & 7 s & 1 m 13 s & 24 m 56 s & -  \\ \hline
        \end{tabular}
        }

        \footnotesize{ 
    \flushleft{  s: seconds, m: minutes, h: hours   \noindent\\
    }
    }
    
    \end{subtable}

\end{table}

Table \ref{Table:scalability} lists the result.
We cited the scalability evaluation results of six clone detectors from two existing studies \cite{SourcererCC, CCAligner}. 
The results for MSCCD and CCAligner are derived from their respective experiments, while the results for the other five tools come from experiments reported in the SourcererCC paper \cite{SourcererCC}.
The machine used for testing MSCCD is comparable to the machines used in these two studies. 
Table \ref{Table:scalability} do not include the result of NIL, SAGA, Oreo,
because the machines they used were not comparable.
It also excludes srcClone because the same experiments were not conducted.
MSCCD was the fastest for 100KLOC to 10MLOC. 
However, MSCCD was slightly slower than the others for 1KLOC to 100KLOC, owing to the overhead of parser execution. 
Also, due to memory requirements, MSCCD could not complete the detection for 100MLOC. 
In such an implementation, MSCCD failed to optimize memory usage sufficiently. 
Considering the situation of SourcererCC, MSCCD has the opportunity to achieve the scalability of 100MLOC through further optimization.


    \noindent\fbox{
        \parbox{0.95\textwidth}{
      The answer to RQ1:
      Compared with existing clone detectors, MSCCD has a median level of recall, precision and scalability.
        }
      }

\section{Multilingually Measuring Recall\&Precision of Code Clone Detectors}
\label{Sec5}

MSCCD supports numerous languages, making evaluating its efficacy across this diverse linguistic landscape crucial. 
To address this need and investigate potential performance variations in the same detection method across different languages, 
we have constructed a benchmark using data from CodeNet\footnote{\url{https://github.com/IBM/Project_CodeNet}}, an online judge System. 
This benchmark serves as a valuable tool to assess the recall and precision in the four languages of the target detector, 
offering insights that extend beyond the constraints of single-language evaluations.

\subsection{Introduction of the benchmark}

Constructing an evaluation benchmark for code clone detectors is an arduous task owing to the requirement of a large code base and marking all true clones in the code pairs.
No clone detector with 100\% recall and precision exists, 
while manual marking of all the actual clones is nearly impossible.
The most significant existing benchmark, BigCloneBench, combines heuristics and manual annotation of many function-level clones \cite{BCB}.
However, it only supports Java. 
Therefore, producing recall--precision benchmarks from real-world code is a near-impossible task.

With the rise of AI for Code research in recent years, 
large-scale annotated datasets are needed for training machine learning models and benchmarking evaluations.
Datasets that contain data from Online Judge systems (OJ systems) play an important role, e.g., CodeNet and CodeXGLUE \cite{CodeNet,CodeXGLUE}.
An OJ system can compile and execute the submitted code and test it to judge whether it is a successful submission for competitive programming.
It can provide information such as source code, task descriptions, execution results, etc.
These data can also evaluate the code clone detector's recall and precision.
All code pairs accepted by the same task can be considered as at least T4 clones. 
The more clones the target detector reports, the higher recall it has.
On the other hand, submissions between tasks that are not identical are not code clones.
Nevertheless, they can also be misreported as clones because they can still contain similar syntacticians (e.g., same methods, variable names, or similar program patterns). 
The more such cases (false positive), the lower the precision of the target detector. 

For this study, we chose CodeNet to implement our ideologies.
CodeNet was proposed by IBM.
It has a collection of over 13,916,000 submissions in 4053 problems from AIZU Online Judge\footnote{\url{https://onlinejudge.u-aizu.ac.jp/home}} and AtCoder\footnote{\url{https://atcoder.jp/}}.
CodeNet contains data from over 50 programming languages. 
While most data originate from C, C++, Java, and Python, 
it is already the largest database of supported languages. 
Besides, identical problems have also been well-labeled.


\begin{figure*}
    \graphicspath{{./img/}}
        \includegraphics[width=0.97\textwidth]{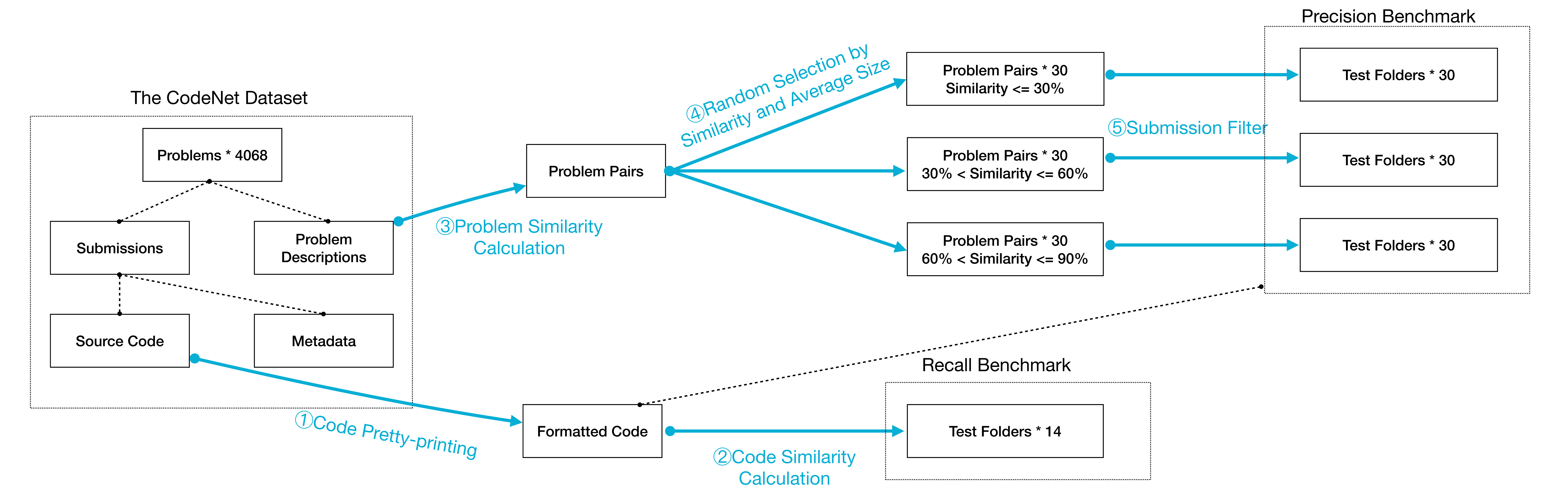}
    \caption{Process of Evaluating Recall\&Precision in Various Languages}
    \label{fig:mainIdea}
\end{figure*}

Fig. \ref{fig:mainIdea} shows the process of generating benchmarks for evaluating recall and precision using data from CodeNet.
We first pre-processed all the source files to remove all comments and contiguous breaks to avoid mismatches of clone matcher (Step 1 in Fig. \ref{fig:mainIdea}).
For evaluating recall, we selected 14 OJ problems from CodeNet and calculated the similarity of all the code pairs accepted by each problem (Step 2). 
The similarity of code pairs from these 14 OJ problems covered all cases from 20\% to 100\%. 
The code submitted to every two problems constituted a subset for evaluating precision. 
We calculated the similarity for every pair of OJ problems by calculating the Jaccard similarity between their descriptive texts (Step 3).
We set up three groups according to the similarity of the problems, 
and each group randomly selected 30 pairs of problems to generate a total of 90 subsets for each language (Steps 4--5).

Using our created benchmark to evaluate code clone detectors,
we used a clone matcher similar to BigCloneBench: line coverage clone matcher \cite{BCB}. 
If the reported clone covered more than a certain percentage of lines of the clones in the benchmark, the clone in the benchmark would be recognized as detected. 
In this study, we used a clone matcher with 70\% coverage.
Hence, we could only determine code clones between files. 
Clone detectors that do not support file-level detection could result in some correct clone reports not being matched by our benchmark. 
Therefore, we used the 70\% configuration, though we believe a higher line coverage threshold should have been used.

We evaluated four detectors: MSCCD, NiCad\footnote{NiCad:\url{https://www.txl.ca/txl-nicaddownload.html}}, CCFinderSW\footnote{CCFinderSW: \url{https://github.com/YuichiSemura/CCFinderSW}}, and SourcererCC\footnote{SourcererCC: \url{https://github.com/Mondego/SourcererCC}}. 
The first three are existing multilingual clone detectors, while SourcererCC also supports C, C++, and Java in the benchmark. 
We used the language support provided by the authors and did not add any changes.
The configurations for each detector are listed in Table \ref{Table:ToolConfigForMultilingualTest}. 
We set the minimal size of tokens to 20 to match our benchmark.
For the other items, we used the default settings of each detector.
Although the research from Ragkhitwetsagul et al. suggests that the default configuration of clone detection tools may not be optimal configuration \cite{ragkhitwetsagul2018comparison}, we opted to use the default settings of each tool in order to facilitate comparison with the evaluation results from existing studies.
Furthermore, in recall measurement in Section \ref{Sec:5.2} and precision measurement in Section \ref{Sec:5.3}, 
we set a limitation for the benchmarks with a minimum token count of 20 and a minimum line count of 6.

\begin{table}[]
\centering
\begin{threeparttable}
\caption{Configurations for the Target Detectors}
\label{Table:ToolConfigForMultilingualTest}
\begin{tabular}{l|l}
\hline
Tool        & Configurations                                                                                                         \\ \hline
MSCCD       & min tokens:20, similarity threshold:70\%, granularity:file                                                                                           \\ \hline
SourcererCC & min tokens:20, similarity threshold:70\%, granularity:file                                                                                               \\ \hline
NiCad       & \begin{tabular}[l]{@{}l@{}}min lines:6, similarity threshold:70\%,\\ blind identifier normalization,\\  identifier abstraction, granularity:file \end{tabular} \\ \hline
CCFinderSW  & min tokens:20                                                                                                          \\ \hline
\end{tabular}



    \end{threeparttable}
\end{table}

\subsection{Benchmarking of Recall}
\label{Sec:5.2}
Existing recall evaluation benchmarks are dedicated to grouping clones according to their type and reporting recall separately for each type of clone \cite{BCB,Svajlenko2021}. 
However, existing clone detectors can automatically distinguish between T1 and T2 only.
Manual annotation of T3 and T4 clones is almost impossible due to the unrealistic workload.

Hence, we attempted to evaluate the recall of a detector from another perspective. 
Numerous similarities can be used to evaluate the similarity between two code segments, including edit distance. 
Edit distance presents the minimum number of operations required to transform one string to another. 
The fewer the changes made to the source code, the more similar the new code segment is to the source code, which fits very well with the process of code clone generation. 
In light of these considerations, 
the proposed benchmark evaluates the coverage of current detection methods for edit distance-based similarity.

\begin{subequations}
    \label{equ:LevTS}

    \begin{equation}
    LevTS(A,B) = 
    \left \{
    \begin{array}{ll}
        |A| & if |B| = 0 \\ 
        |B| & if |A| = 0 \\ 
        LevTS(tail(A), tail(B)) & A[0] = B[0] \\
        1 + min \left \{
            \begin{array}{l}
                LevTS(tail(A),B) \\
                LevTS(A, tail(B)) \\
                LevTS(tail(A), tail(B))
            \end{array}
        \right.  & otherwise
    \end{array}
    \right. 
\end{equation}
\begin{equation}
    LevTS(A,B) \epsilon \left [ 0,   max\left ( \left | A \right | , \left | B \right | \right )   \right ]
\end{equation}
\end{subequations}

\begin{equation}
    \label{equ:LevSimi}
    LevSimi(A,B) = \frac{Max\left ( \left | A \right |,  \left | B \right |\right ) - LevTS\left (A,B\right )}{Max\left ( \left | A \right |,  \left | B \right |\right )} , LevSimi(A,B)  \epsilon \left [ 0,1 \right ]
\end{equation}

We used the Levenshtein distance, 
an edit distance that allows three operations: insertion, deletion, or substitution \cite{10.1145/375360.375365}. 
We applied the distance to token sequences instead of strings.
The Levenshtein distance $LevSimi_{(A,B)}$ of two token sequences $A$ and $B$ was calculated using Eq. \ref{equ:LevTS}. 
The edit distance similarity between the two token sequences $A$ and $B$ was defined in Eq. \ref{equ:LevSimi}, such that one minus the edit distance of $A$ and $B$ was divided by the length of the more extensive token sequence.
In addition, most modern code clone detectors can accurately detect T1 and T2 clones.
Therefore, we normalized all identifiers as the same token before computation. 
Thus, T1 and T2 clones were both computed with a 100\% similarity.

\begin{figure}
    \centering
    \graphicspath{{./img/}}
    \includegraphics[width=0.99\textwidth]{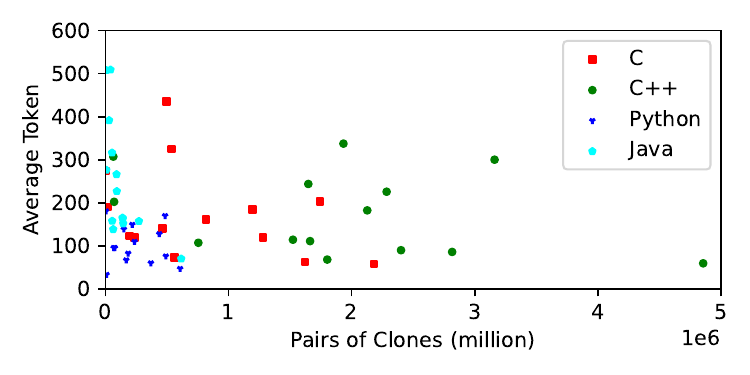}
    \caption{Distribution of Four Target Languages in Pair Number and Average Token}
    \label{fig:recall_simi_dis}
\end{figure}

We calculated the similarity of all code pairs in all successful submissions for the first 2,300 problems and counted their distribution from 0.20 to 1.00. 
Thereafter, we chose 14 problems with the full distribution of LevTs $\Theta$ in the collection $\left \{0.20 \leq \Theta \leq 1.00 \mid  \Theta  mod  0.01 = 0   \right \}$ for all targeted languages. 
Fig. \ref{fig:recall_simi_dis} shows the distribution of the 14 problems across a total of 56 sub-datasets in the four target languages for the number of code pairs and the average number of tokens. 
C++ had the highest number of files, while
Java had relatively longer programs. Python had significantly fewer average tokens than the other languages.

For each evaluation object (detector-language), we performed clone detection on each of the 14 sub-datasets, 
and then the final results were aggregated. 
We reported the recall for each group in groups at 0.20 intervals, which are presented in Table \ref{Table:Recall_CodeNet}. 
The parts with 100\% similarity were T1 and T2 clones,
while the other corresponded to T4 and T3 clones.
For areas of T1 and T2, CCFinderSW (famous for detecting T1 and T2 clones) ranked first for Java and Python.
CCFinderSW also got a near-perfect result of 98\% for C and ranked 2nd behind NiCad in this area.
MSCCD ranks second in C++, but the gap between it and the top-ranked SourcererCC is extremely small, 
with less than a 1\% difference in the number of detected clones.
In the T3 and T4 areas, the recall of all four tools began to show a significant decline.
For the LevSimi interval between 80\% to 99\%, 
MSCCD and SourcererCC ranked in the top two in the three languages other than Python, while NiCad ranked first in Python.
For the intervals under 79\% similarity, none of the detectors had a recall of more than 40\% for all languages, 
while MSCCD ranked 1st for the most time in these intervals.

For the same detectors, performance differences between languages were also evident. 
For example, during the interval [0.8,0.99], MSCCD performed best for Java (81\%), followed by C++ (73\%) and C (70\%), and was relatively weakest for Python (42\%). 
NiCad also had a significant difference in this interval, with a 49\% recall for Python and C, while the recall for Java is 15\%.
SourcererCC showed results that were similar to MSCCD because we believe that as statically strongly typed languages, Java, C, and C++ have a higher rate of keyword occurrences in their syntax, such as the description of classes, functions, and types of variables, that assist the detector in identifying clones. 
Since Java naturally shares more tokens for similar programs and thus has a higher degree of similarity because it has a fixed class, differences between Java and C/C++ were observed. 
Contrarily, C/C++ source programs contain more pre-processing commands, which brings more dissimilarity.


A noteworthy observation in the results of CCFinderSW was that the C++ trials of CCFinderSW had only 84\% recall at the Type-1\&2 area. 
This may have been due to the high frequency of the standard input--output stream in the OJ system. 
For example, there was no significant difference in the percentage of using C standard statements, such as the printf method from ``stdio.h,’’ and C++ standard statements, such as the cout method from ``iostream’’ among C++ programs in CodeNet.
This led to several false negatives in CCFinderSW based on token sequence detection of clones.


\begin{table}
    \centering
    \caption{Recall Measurement Grouped by LevSimi Range}
    \label{Table:Recall_CodeNet}
    \resizebox{\textwidth}{!}{
    \begin{tabular}{c|c|c|c|c|c|c|c|c|c|c|c}
        \hline
        \multirow{4}{*}{Language} & \multirow{4}{*}{Tool} & \multicolumn{10}{c}{Recall for Each Similarity Range}  \\ \cline{3-12}
         & & \multicolumn{8}{c|}{T3 or T4} & \multicolumn{2}{c}{T1 or T2} \\ \cline{3-12}
         & & \multicolumn{2}{c|}{[0.20, 0.39]} & \multicolumn{2}{c|}{[0.40, 0.59]} & \multicolumn{2}{c|}{[0.60, 0.79]} & \multicolumn{2}{c|}{[0.80, 0.99]}  & \multicolumn{2}{c}{1.0} \\ \cline{3-12}
         & & \% & \# &\% & \# & \% & \# & \% & \# & \% & \# \\ \hline 
         \multirow{3}{*}{Java}& MSCCD       & \textbf{1} & 1,940  & 2 &13,353  & 28  &101,345  & 81 & 275,952  & 90 & 16,417\\
                              & SourcererCC & 1 & 1,905  & \textbf{2} & 15,581 & \textbf{29} & 102,499 &  \textbf{82} & 278,953  & 90 & 16,281 \\
                              & NiCad       & 0 & 0  & 0 & 16 & 1 & 1,969 &  15 & 52,507  & 99 & 18,000 \\
                              & CCFinderSW  & 0 & 0 & 0 & 0 & 0 & 1,188 & 22 & 76,052  & \textbf{100} & 18,163 \\ \hline
        \multirow{3}{*}{Python}& MSCCD       & \textbf{0} & 792 & \textbf{1} & 16,106 & \textbf{9} & 39,906 & 42 & 26,046  & 79 & 2,776 \\
                              & SourcererCC & - & - & - & - & - & - & - & -  & - & - \\
                              & NiCad & 0 & 2  & 0 & 768 & 2 & 10,053 &  \textbf{49} & 30,303  & 98 & 3,447 \\
                              & CCFinderSW  & 0 & 3 & 0 & 310 & 1 & 4,417 & 21 & 13,184  & \textbf{99} & 3,492 \\ \hline
        \multirow{3}{*}{C}& MSCCD       & \textbf{8} & 153,559 & \textbf{10} & 419,469 & \textbf{37} & 844,211 & 70  & 1,455,265  & 78 & 107,235 \\
                          & SourcererCC & 7 & 140,581 & 9 & 392,206 & 37 & 837,573  & \textbf{70} & 1,462,909 & 78 & 107,361 \\
                          & NiCad & 0 & 0  & 0 & 669 & 3 & 77,009 &  49 & 1,029,109  & \textbf{100} & 136,913 \\
                          & CCFinderSW  & 0 & 0 & 0 & 4 & 0 & 9,118 & 17 & 346,114  & 98 & 134,722 \\ \hline
        \multirow{3}{*}{C++}& MSCCD       & \textbf{0} & 16,756 & \textbf{3} & 345,554 & \textbf{34} & 1,466,095 & \textbf{73} & 1,415,698  & 90 & 75,024\\
                            & SourcererCC & 0 & 10,897 & 2 & 216,535 & 28 & 1,208,668 & 71 & 1,383,606  & \textbf{90} & 75,609 \\
                            & NiCad & - & -  & - & - & - & - &  - & -  & - & - \\
                            & CCFinderSW  & 0 & 0 & 0 & 53 & 0 & 1,188 & 6 & 124,825  & 84 & 70,275 \\ \hline
        
    \end{tabular}}

\footnotesize{ 
\flushleft{  \%: Recall in precentage, \#: Number of clones detected   \noindent\\
}
}
    
\end{table}

\begin{sidewaysfigure}
    \begin{subfigure}{\textwidth}
        \includegraphics*[width=0.55\textwidth, height=0.3\textheight]{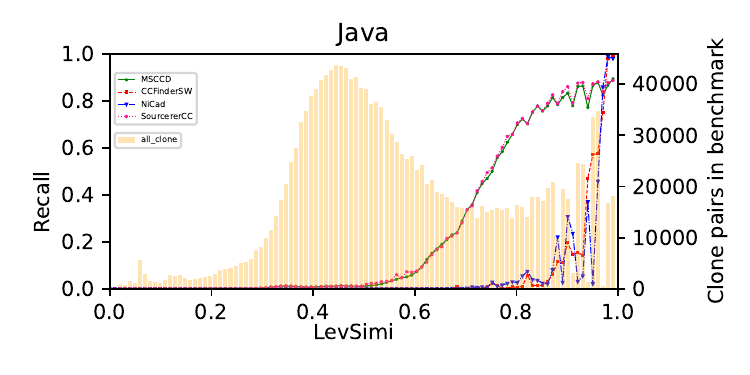}
        \includegraphics*[width=0.55\textwidth, height=0.3\textheight]{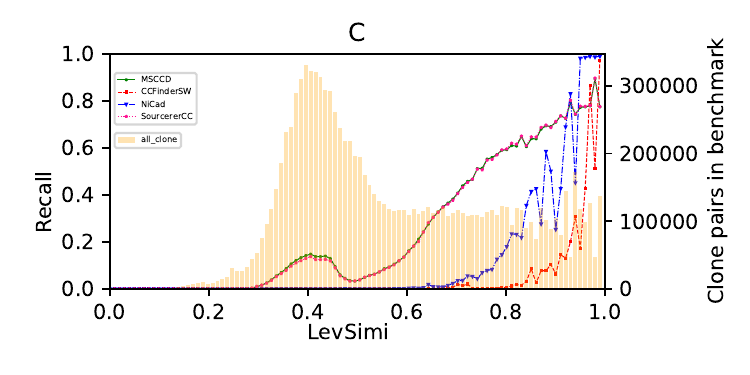}
    \end{subfigure}
    \begin{subfigure}{\textwidth}
       \includegraphics*[width=0.55\textwidth, height=0.3\textheight]{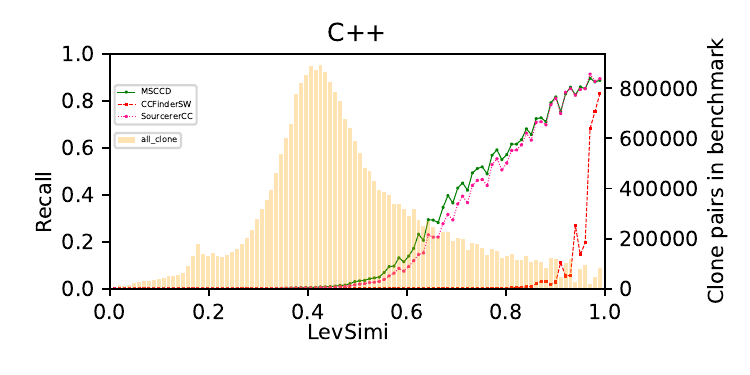}
       \includegraphics*[width=0.55\textwidth, height=0.3\textheight]{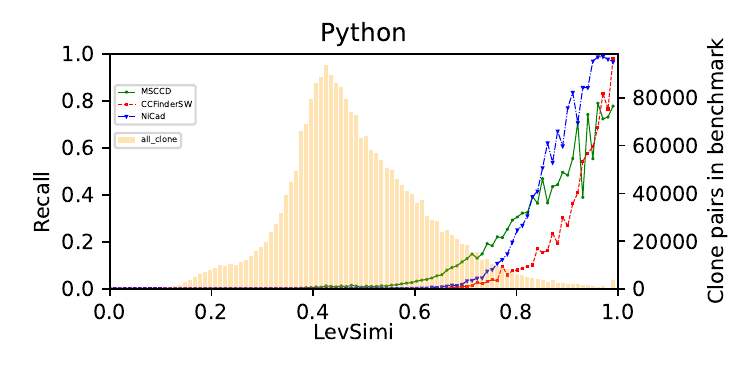}
    \end{subfigure}

    \caption{Recall at Each LevSimi}
    \label{fig:Recall_CodeNet}

\end{sidewaysfigure}

For a better visual representation of the degree of coverage of edit distance-based similarity by each detection method, 
we plotted a line graph, as shown in Fig. \ref{fig:Recall_CodeNet}. 
The horizontal axis represents the edit distance similarity, 
and the yellow bars represent the number of clones included in the benchmark under that similarity. 
The higher the number of clones, the more likely the evaluation results at that location would have increased the confidence. 
Moreover, the dots on the dash represent the recall of the tool at that edit distance similarity.
We observed that MSCCD and SourcerCC had the same performance and had a relatively higher reproducibility in Type-3 detection compared to NiCad and CCFinderSW. 
Large fluctuations in individual similarities (even at highly similar positions) were observed because the similarities used by both CCFinderSW and NiCad were very sensitive to changes at the statement level, 
especially when the changes were located in the middle of the program. 
Moreover, such types of code clones were unevenly distributed in the benchmark, 
thus producing significant fluctuations.

\subsection{Benchmarking of Precision}
\label{Sec:5.3}
\newcommand{\RomanNumeralCaps}[1]
    {\MakeUppercase{\romannumeral #1}}

For precision evaluations, we constructed a benchmark based on ``true positives if two blocks of clones are detected from the same problem, and false positives if they are from different problems.’’ 
The basic process of the experiment was to randomly select a certain number of problems from more than 4000 problems included in CodeNet and gather all the source files into a folder. 
Thereafter, each target detector was used to detect clones of each folder.
The detected clones were considered true positives if they were from the same problem or false positives in case they were from different problems.
The use of meaningless variable names, such as \verb|i| and \verb|j|, 
and some common methods, such as reading from the standard input or printing to the standard output, were widespread in programs submitted to the OJ system. 
Thus, code pairs from different problems still had the potential to be mistakenly reported as code clones by the clone detector. 
Thus, a challenge that affected the correctness of the clone detector was observed.
The experiment evaluated how the code clone detectors coped with such a situation.

We used the Jaccard similarity to calculate the similarity between the descriptive texts of the problems. 
The higher the similarity between problems, the more likely it is for the code from these problems to be falsely reported as code clones. 
In other words, this pair of problems represents a more challenging case for precision testing.
Only a part of the problem statement was involved in the computation for each problem. 
Constraints, input, output, and sample were not computed.
We randomly sampled three groups with the similarity $\sigma$ of $\sigma\in [0,0.3)$ (group \RomanNumeralCaps{1}), $\sigma\in [0.3,0.6)$ (group \RomanNumeralCaps{2}), and $\sigma\in [0.6,0.9)$ (group \RomanNumeralCaps{3}), 
and each \RomanNumeralCaps{3} consisted of 30 pairs of problems. 
Overlap was not allowed between pairs in the same group.
In addition, the difference in the average code length per pair of problems did not exceed 30\%.

The results are presented in Table \ref{Table:precisionCodeNet}.
All target tools were performed with near-perfect precision for groups \RomanNumeralCaps{1} and \RomanNumeralCaps{2}.
The similarity of the problems in these two groups was less than 60\%. 
Clone detectors worked well even when the same identifier name or function was applied. 
The differences between code clone detectors and languages mainly appeared in group \RomanNumeralCaps{3}.
The problem pairs in group \RomanNumeralCaps{3} were more similar to groups \RomanNumeralCaps{1} and \RomanNumeralCaps{2}, and most were modified versions of existing problems. 
For example, problems p01950 and p01964 were aimed at implementing a breadth--firth search.
P01964 was based on a directed graph, while p01950 was based on an undirected graph. 
There were many submissions for this class of problems based on modifications of the previous submissions. 
The changes were usually the addition of a function and a shift in the main function. 
The class constituted some large-gap clones, although they were considered non-clones by benchmark because they were from a different problem.

The results for MSCCD and SourcererCC were very close, although SourcererCC did not support Python, 
and both achieved relatively high precisions in Python and C++ (above 85\%), 
followed by Java (slightly below 80\%), 
and the lowest precision for C (slightly below 65\%). 
For the class ``large-gap-clone-like clone’’ mentioned previously, 
the detection algorithm of MSCCD and SourcererCC were likely to be reported as code clones. 

\begin{table}

    \centering
    \caption{Precision Measurements Grouped by Similarity of Problem Description }
    \label{Table:precisionCodeNet}
    \resizebox{0.8\textwidth}{!}{
    \begin{tabular}{c|c|c|c|c}
        \hline
        \multirow{2}{*}{Language} & \multirow{2}{*}{Tool} & \multicolumn{3}{c}{Precision (median) / \% }  \\ \cline{3-5}
         & &  \specialcell{Group \RomanNumeralCaps{1} \\ (0,0.3)}&\specialcell{Group \RomanNumeralCaps{2} \\ (0.3,0.6)} & \specialcell{Group  \RomanNumeralCaps{3} \\ (0.6,0.9)}  \\ \hline
        \multirow{3}{*}{Java} & MSCCD            & 99.7(100) & 96.8(100) & 78.0(67.7) \\ 
                              & SourcererCC      & 98.5(100) & 97.0(100) & 78.9(64)   \\
                              & CCFinderSW       & 99.8(100) & 94.5(100) & 99.9(80)   \\ 
                              & NiCad            & 100(100)  & 100(100)  & 95.1(80.8) \\\hline
        \multirow{3}{*}{Python} & MSCCD       & 99.9(100) & 99.8(100) & 86.7(83.3) \\
                                & SourcererCC & - & - & -  \\
                                & CCFinderSW  &  99.9(100) & 99.9(100)  & 99.2(100)   \\ 
                                & NiCad         & 100(100)  & 100(100)  & 99.9(100) \\ \hline
        \multirow{3}{*}{C} & MSCCD        & 100(100) & 99.6(100) & 64.1(57.0)  \\
                           & SourcererCC  & 99.7(100) & 99.7(100) & 64.3(57.1) \\
                           & CCFinderSW   & 100(100) & 99.9(100) & 99.9(99.9)  \\ 
                              & NiCad       & 100(100)  & 99.9(100) & 70.1(100) \\ \hline
        \multirow{3}{*}{C++} & MSCCD        & 99.8(100) & 99.8(99.9) & 86.6(68.1) \\
                             & SourcererCC  & 99.4(100) & 91.0(99.9) & 86.5(60.7) \\
                             & CCFinderSW   & 99.9(100)  & 99.5(100)  & 70.1(66.7) \\  
                              & NiCad       & -  & - & -  \\\hline

    \end{tabular}
    }
\end{table}

Conversely, CCFinderSW still achieved near-perfect precisions in Java, Python, and C. 
However, since CCFinderSW cannot detect Type-3 code clones, the parts it reported were relatively less likely to be false positives because of their higher syntactic similarity. 
Since the detection algorithm of CCFinderSW is susceptible to statement insertion, many false positives were reduced when compared to MSCCD and SourcererCC.
The exception was for C++, where the precision of CCFinderSW (70.1\%) was significantly lower than that of MSCCD and SourcererCC, which was related to the clone-matcher setting used in the benchmark. 
In the CodeNet dataset, 
C++ programs more commonly had longer pre-process commands or macro definitions. 
Hence, even though there were statement-level differences in the functions, many \#include, and \#define statements were matched, allowing many results to be matched by the clone matcher and counted as false positives.

NiCad had results similar to CCFinderSW for Java and Python, but the precision was decreased for C (70\%). 
MSCCD and SourcererCC also had the lowest precision in C, which may have been due to the more significant proportion of indistinguishable clones among the 30 pairs of problems in the C language of the group \RomanNumeralCaps{3}. 
The similarity of the problem description texts and the average size of the submissions were considered in extracting the problem pairs. Hence, the selection of the same problem pairs for all four languages was impossible. 
To avoid this effect, we also recorded the median results in each group. 
In contrast to the general case, the median precision for the C language of NiCad was higher than its ensemble value. 
Thus, the presence of pairs with more submissions with a higher percentage of indistinguishable clones was observed, which affected the ensemble results.

\noindent\fbox{
  \parbox{0.95\textwidth}{
The answer to RQ2:
All the evaluated detectors showed different performances between languages in recall and precision, and the difference in performance between languages in recall was more pronounced than precision. Users should try different settings for different languages (mainly including similarity thresholds) to achieve more accurate detection results.}
  }

\section{Language Extensibility of Existing Multilingual Code Clone Detectors}

\label{Sec6}

This section discusses the language extensibility of the existing multilingual clone detectors. 
Any code clone detection approach can be applied to any language with sufficient effort and time. 
Owing to the development costs and required knowledge, 
most code clone detectors have not been extended to support any new languages since their first release. 
Therefore, we define the language extensibility of a multilingual clone detector for a number of languages that can be supported without having to write or modify the program. 
We count the number of languages supported by these tools.
Moreover, we discuss the language extensibility of MSCCD, CCFinderSW, and NiCad, considering the conditions for supporting a new language, the difficulty, the languages supported by the current version, and the available grammar definition resources.
The results are listed in Table \ref{Table:LanguaageExtensibilityCompare}.

\subsection{MSCCD}
MSCCD showed conditional support for the targeted language based on the correctness of the ANTLR4 grammar definition file. 
If an incorrect grammar definition file is used (e.g., due to a version mismatch or inclusion error), 
it could result in the failure of the parser generation, trigger an error in parsing, and lead to an unsuccessful creation of an accurate and complete parse tree (PT).
MSCCD could only complete file-level clone detection when the most severe failure occurred.
In addition, MSCCD contains the keywords filter, which can be used to speed up the detection by reducing extracted token bags.
When the grammar of object composition of the target language did not have the features utilized by the keywords filter, that composition would not be extracted when the filter was activated.

\subsubsection{Language Extensibility Measurement by Token Bag Generation}
\label{Sec6.1.1}
Similar to previous experiments in our initial paper \cite{MSCCD}, 
we randomly selected programs from Rosetta Code\footnote{\url{http://www.rosettacode.org/wiki/Rosetta_Code}} to check whether MSCCD could generate the correct token bag for all the target compositions of the targeted language. 
Rosetta Code is a programming chrestomathy site that collects instances of multiple languages for over a thousand programming tasks.
The site primarily shows case programs that are implemented practically and simply without relying heavily on development frameworks.
This ensures that the extracted programs contain the target compositions.
The target languages for the experiment were those supported by RosettaCode and Grammars-v4 for 48 languages. 
We first attempted to generate the MSCCD token bag generator, which includes a specific parser for each language.
We randomly selected five programs from Rosetta Code for the languages whose token bag extractor was generated successfully.
After that, we checked whether they could generate the correct token bag with or without the keywords filter in effect.

Regarding the keywords filter activated,
we selected four compositions for each language: class, function, branch, and loop. 
These items are the most basic and standard parts of languages. 
For each item, all corresponding grammar was included within the scope of the test targets. 
For example, all loop items were test objects, including but not limited to the while loop, for loop, and do-while loop. 
We regarded function items as the most relevant to show the target language's extensibility among the four items. 
The other three items indicated whether MSCCD could detect code clones at higher granularity. 
However, MSCCD did not merely support these four items. 
The other language components could also be extracted depending on the grammar. 
The item was passed when all corresponding code blocks were extracted correctly, 
including the corresponding code block, lacking any irrelevant code segments (circle mark; \CIRCLE). 
It was regarded as a failure when a target code block failed to be generated correctly without a syntax analysis error \cite{MSCCD}.

\begin{table}
    \caption{Language Extensibility of the Existing Tools}
    \label{Table:LanguaageExtensibilityCompare}
    \resizebox{\textwidth}{!}{

\begin{tabular}{c|c|c}
\hline
Tool       & \begin{tabular}[c]{@{}c@{}}Conditions for \\ Supporting the Language\end{tabular}                                  & \begin{tabular}[c]{@{}c@{}}Positive Rate of\\ Evaluation Experime\end{tabular}             \\ \hline
MSCCD      & Available ANTLR grammar definition files                                                                           & \begin{tabular}[c]{@{}c@{}}89.58\% (43 / 48)\\ (18 languages file-level only)\end{tabular} \\ \hline
CCFinderSW & \begin{tabular}[c]{@{}c@{}}Grammar of comment and string \\ can be converted into regular expressions\end{tabular} & 84.62\% (44/52)                                                                            \\ \hline
NiCad      & Available TXL grammar definition files                                                                             & -                                                                                          \\ \hline  \hline
Tool       & \begin{tabular}[c]{@{}c@{}}Number of Languages \\ Currently Supported\end{tabular}                                 & Available Resources                                                                        \\ \hline
MSCCD      & 25                                                                                                                 & more than 100                                                                              \\ \hline
CCFinderSW & 15                                                                                                                 & more than 100                                                                              \\ \hline
NiCad      & 10                                                                                                                  & 22                                                                                         \\ \hline
\end{tabular}
    }
\end{table}

When the keywords filter was unactivated, all compositions were extracted when the PT was generated successfully. 
Therefore, the results of individual components were not distinguished in this case.
We called it a block/statement level. 
A positive result (circle mark; \CIRCLE) meant that the ANTLR-generated parser generated the correct PT for all test source files.

\begin{table}
    \centering
    \caption{Measurements of MSCCD's Language Extensibility}
    \label{Table:LanguageExtensibility}
    \resizebox{\textwidth}{!}{
    \begin{tabular}{r|cccc|c}
        \hline
        \hline

        \multicolumn{2}{c|}{Support Status}         & \multicolumn{2}{c|}{Language Number}  &  \multicolumn{2}{c}{Notes} \\\hline \hline
        \multicolumn{2}{c|}{block-level     }& \multicolumn{2}{c|}{25 }& \multicolumn{2}{c}{-}\\ \hline
        \multicolumn{2}{c|}{file-level only }& \multicolumn{2}{c|}{18 }& \multicolumn{2}{c}{error in syntactical analysis }\\ \hline
        \multicolumn{2}{c|}{not supported   }& \multicolumn{2}{c|}{5  }& \multicolumn{2}{c}{\specialcell{cursh in parser generation,\\ severer error in lexical analysis}} \\         
        
        \hline
        \hline

        \multirow{3}{*}{Language} & \multicolumn{5}{c}{Results of 25 Block-level Supported Languages} \\ \cline{2-6}
                   & \multicolumn{4}{c|}{Keywords Filter Activated} & Keywords Filter Inactivated \\ \cline{2-6}
                   & Class & Function & Condition & Loop & Block/Statement \\ \hline
        C          & \CIRCLE  & \CIRCLE     & \CIRCLE      & \CIRCLE & \CIRCLE       \\
        C\#        & \CIRCLE  & \CIRCLE     & \CIRCLE      & \CIRCLE & \CIRCLE       \\
        C++        & \hexstar & \CIRCLE     & \CIRCLE      & \CIRCLE & \CIRCLE       \\
        Cobol85    & -        & \CIRCLE     & \CIRCLE      & \CIRCLE & \CIRCLE       \\
        Clojure    & -        & -           & -            & -       & \CIRCLE       \\
        Cmake      & -        & -           & -            & -       & \CIRCLE       \\
        Dart       & \hexstar & \CIRCLE     & \CIRCLE      & \CIRCLE & \CIRCLE       \\
        Erlang     & -        & -           & \CIRCLE      & \CIRCLE & \CIRCLE       \\
        Go         & -        & \CIRCLE     & \CIRCLE      & \CIRCLE & \CIRCLE       \\
        Java       & \CIRCLE  & \CIRCLE     & \CIRCLE      & \CIRCLE & \CIRCLE       \\
        JavaScript & \CIRCLE  & \CIRCLE     & \CIRCLE      & \CIRCLE & \CIRCLE       \\
        Kotlin     & \CIRCLE  & \CIRCLE     & \CIRCLE      & \CIRCLE & \CIRCLE       \\
        Lua        & \CIRCLE  & \CIRCLE     & \CIRCLE      & \CIRCLE & \CIRCLE       \\
        Objective-C& -        & \hexstar    & \CIRCLE      & \CIRCLE & \CIRCLE       \\
        PHP        & \CIRCLE  & \CIRCLE     & \CIRCLE      & \CIRCLE & \CIRCLE       \\
        prolog     & -        & -           & -            & -       & \CIRCLE       \\
        Python     & \CIRCLE  & \CIRCLE     & \CIRCLE      & \CIRCLE & \CIRCLE       \\
        Rexx       & -        & -           & -            & -       & \CIRCLE       \\
        Rust       & \CIRCLE  & \CIRCLE     & \CIRCLE      & \CIRCLE & \CIRCLE       \\
        Swift      & \CIRCLE  & \CIRCLE     & \CIRCLE      & \CIRCLE & \CIRCLE       \\
        TypeScript & \CIRCLE  & \CIRCLE     & \CIRCLE      & \CIRCLE & \CIRCLE       \\
        VBA        & -        & \CIRCLE     & \CIRCLE      & \CIRCLE & \CIRCLE       \\
        Verolog    & -        & -           & -            & -       & \CIRCLE       \\
        VHDL       & -        & -           & -            & -       & \CIRCLE       \\
        VB         & -        & \CIRCLE     & \CIRCLE      & \CIRCLE & \CIRCLE       \\ \hline 
        
    \end{tabular}
    }
    \footnotesize{\flushleft{ \CIRCLE: Positive   \hexstar: Negative  -: Not applicable \noindent\\
}}
\end{table}

Out of the 48 targeted languages, MSCCD did not support five languages: 
Three failed during parser generation, and two encountered errors during lexical analysis. 
An additional 18 languages encountered severe syntax analysis errors, and therefore, 
MSCCD could only perform file-level detection for these. 
The detailed results for the successfully processed 25 languages are listed in Table \ref{Table:LanguageExtensibility}.
When the keywords filter was turned off, all 25 languages were supported. 
Dart, C++ classes, and objective-C functions were not fully extracted when the filter was used because they used a different pattern in their syntax that the filter could not recognize. 
The remaining components were extracted successfully.

\subsubsection{Details of ANTLR Parser Errors}

In the experiment of Section \ref{Sec6.1.1}, some target languages were judged as not supporting the block/statement level due to syntactic analysis errors.
In fact, not all errors directly affect MSCCD's support for the language.
Different types and locations of errors have varying degrees of impact.
If the main part of the parse tree in successfully constructed, 
syntactic errors may not affect the outcome of token bag extraction.

To evaluate the impact of ANTLR parser errors on token bag generation in practical use,
we conducted an experiment targeting open-source software (OSS). 
We selected five languages from grammars-v4 that were explicitly marked with language versions and chose an OSS from GitHub based on stars, which match the language version, 
and containing more than 50 source files.
The details of selected OSS are listed in Table \ref{Table:ParserErrorDetails.a}.
The minimum token size of MSCCD was set to 30.
We collected parser error messages for each source file and tallied the number of token bags generated under various error statuses. 

\begin{table}[h]
    \centering
    \caption{Generated Token Bags in Each Parser Error Situation}
    \label{Table:ParserErrorDetails}
    \begin{subtable}[t]{\textwidth}
    \centering
    \caption{Details of Targeted Open-Source Software}
    \label{Table:ParserErrorDetails.a}
            \resizebox{\textwidth}{!}{
        \begin{tabular}{c|c|ccc}
        \hline
          \multirow{2}{*}{Language} & \multirow{2}{*}{Version} & \multicolumn{3}{c}{OSS}  \\ \cline{3-5}
                                    &                          & \multicolumn{1}{c|}{Name}  & \multicolumn{1}{c|}{Files} & LOC   \\ \hline
           Java                      & Java20                   & \multicolumn{1}{c|}{Spring-boot-3.1.3}         & \multicolumn{1}{c|}{6888}  & 679,651 \\
           JavaScript                & ECMAScript 2020          & \multicolumn{1}{c|}{react-18.2.0}              & \multicolumn{1}{c|}{1882}  & 450,759\\
           C\#                       & CSharp6                  & \multicolumn{1}{c|}{shadowsocks-windows-3.4.3} & \multicolumn{1}{c|}{127}   & 25,099\\
           C++                       & CPP14                    & \multicolumn{1}{c|}{tensorflow-1.4.0}          & \multicolumn{1}{c|}{2314}  & 621,573\\
           Python                    & Python3.6                & \multicolumn{1}{c|}{django-2.0.5}              & \multicolumn{1}{c|}{2454}  & 320,912\\ \hline
        \end{tabular}}
    \end{subtable}
    
    \hfill

    \begin{subtable}[t]{\textwidth}
    \caption{File Numbers and Generated Token Bags}
    \label{Table:ParserErrorDetails.b}
    \begin{threeparttable}
    \resizebox{\textwidth}{!}{
        \begin{tabular}{c|cccccccc|cccc}
        \hline
        \multirow{3}{*}{Language} & \multicolumn{8}{c|}{Error Type} & \multicolumn{4}{c}{Error Position} \\ \cline{2-13}
        & \multicolumn{2}{c|}{No Error}                          & \multicolumn{2}{c|}{$SE$ Only \tnote{a}}                 & \multicolumn{2}{c|}{$LE$ Only \tnote{b}}                   & \multicolumn{2}{c|}{$LE$ \& $SE$}   & \multicolumn{2}{c|}{Head \tnote{c}}                             & \multicolumn{2}{c}{Tail \tnote{d}}          \\ \cline{2-13}
        &  \multicolumn{1}{c|}{Files} & \multicolumn{1}{c|}{$n$}    & \multicolumn{1}{c|}{Files} & \multicolumn{1}{c|}{$n$}    & \multicolumn{1}{c|}{Files} & \multicolumn{1}{c|}{$n$}     & \multicolumn{1}{c|}{Files} & $n$    & \multicolumn{1}{c|}{Files} & \multicolumn{1}{c|}{$n$}    & \multicolumn{1}{c|}{Files} & $n$   \\ \cline{1-13}
        Java & \multicolumn{1}{c|}{6659}  & \multicolumn{1}{c|}{6.38} & \multicolumn{1}{c|}{217}   & \multicolumn{1}{c|}{6.33} & \multicolumn{1}{c|}{0}     & \multicolumn{1}{c|}{-}     & \multicolumn{1}{c|}{0}     & -    & \multicolumn{1}{c|}{34}    & \multicolumn{1}{c|}{6.56} & \multicolumn{1}{c|}{46}    & 4.52 \\
        JavaScript & \multicolumn{1}{c|}{782}   & \multicolumn{1}{c|}{9.54} & \multicolumn{1}{c|}{1100}  & \multicolumn{1}{c|}{2.21} & \multicolumn{1}{c|}{0}     & \multicolumn{1}{c|}{-}     & \multicolumn{1}{c|}{0}     & -    & \multicolumn{1}{c|}{651}   & \multicolumn{1}{c|}{1.69} & \multicolumn{1}{c|}{63}    & 6.13 \\
        C\#  & \multicolumn{1}{c|}{51}    & \multicolumn{1}{c|}{8.90} & \multicolumn{1}{c|}{0}     & \multicolumn{1}{c|}{-} & \multicolumn{1}{c|}{76}    & \multicolumn{1}{c|}{10.01} & \multicolumn{1}{c|}{0}    & - & \multicolumn{1}{c|}{76}    & \multicolumn{1}{c|}{10.01} & \multicolumn{1}{c|}{0}     & -    \\
        C++ & \multicolumn{1}{c|}{1861}  & \multicolumn{1}{c|}{9.46} & \multicolumn{1}{c|}{388}   & \multicolumn{1}{c|}{7.73} & \multicolumn{1}{c|}{0}     & \multicolumn{1}{c|}{-}     & \multicolumn{1}{c|}{65}    & 4.39 & \multicolumn{1}{c|}{77}    & \multicolumn{1}{c|}{5.76} & \multicolumn{1}{c|}{91}    & 9.31 \\
        Python & \multicolumn{1}{c|}{2452}  & \multicolumn{1}{c|}{8.55} & \multicolumn{1}{c|}{2}     & \multicolumn{1}{c|}{8.59} & \multicolumn{1}{c|}{0}     & \multicolumn{1}{c|}{-}     & \multicolumn{1}{c|}{0}     & -    & \multicolumn{1}{c|}{0}     & \multicolumn{1}{c|}{-}    & \multicolumn{1}{c|}{0}     & -    \\ \hline
        \end{tabular}}

            

            \begin{tablenotes}
        \tiny
        \begin{minipage}[t]{0.5\textwidth}
            \item[a] $SE$: Syntax Error
        \end{minipage}%
        \begin{minipage}[t]{0.5\textwidth}
            \item[b] $LE$: Lexel Error
        \end{minipage}
        
        \begin{minipage}[t]{0.5\textwidth}
            \item[c] Head: The first error occurs at the first 20\% lines
        \end{minipage}%
        \begin{minipage}[t]{0.5\textwidth}
            \item[d] Tail: The first error occurs at the last 20\% lines.
        \end{minipage}
        \begin{minipage}[t]{0.5\textwidth}
            \item[e] $n$: token bags generated for every 100 lines of code
        \end{minipage}
    \end{tablenotes}
    \end{threeparttable}
    \end{subtable}
\end{table}

Table \ref{Table:ParserErrorDetails.b} lists the number of files and generated token bags in each parse error situation.

Python exhibited almost no errors.
Java and C++ were relatively insensitive to syntactic analysis errors, 
with no significant decrease in the number of token bags generated when such errors occurred.
JavaScript was the most sensitive to syntactic errors: 
The number of token bags generated per 100 LOC in the presence of syntactic errors was less than a quarter of that generated without errors.
The situation with C\# was unique, as it produced many lexical errors and no syntactic errors.
Upon investigation, these lexical errors were all due to Byte Order Mark (BOM) symbols in the source files, 
which the lexical analyzer failed to recognize correctly. 
However, this did not affect the syntactic analysis.

For C++, when lexical and syntactic errors coincided, the average number of token bags decreased from 7.73, when only syntactic errors were present, to 4.39. 
This indicates that the impact of lexical errors is more significant than syntactic errors. 
The earlier the occurrence of either syntactic or lexical errors, the greater their impact, a phenomenon that was reflected in the results for both JavaScript and C++.

\begin{table}[]
\centering
\caption{Proportion of Error Messages}
\label{Table:ProportionOfErrorMessages}
\begin{tabular}{c|c|c}
\hline
Message Type                       & JavaScript & C++     \\ \hline
no viable alternative at input ... & 63.46\%    & 8.74\%  \\ \hline
mismatched input ...               & 18.25\%    & 59.11\% \\ \hline
extraneous input ...               & 15.78\%    & 4.21\%  \\ \hline
missing ...                        & 2.27\%     & 27.93\% \\ \hline
others                             & 0.23\%     & 0       \\ \hline
\end{tabular}
\end{table}

\begin{table}[]
\caption{The 5 Most Frequent Error Messages of JavaScript and C++}
\label{Table:TopParserErrorMessages}
\resizebox{\textwidth}{!}{
\begin{tabular}{c|cc|cc}
\hline
\multirow{2}{*}{Ranking} & \multicolumn{2}{c|}{JavaScript}                                        & \multicolumn{2}{c}{C++}                                   \\ \cline{2-5} 
                         & \multicolumn{1}{c|}{Message}                             & Persentages & \multicolumn{1}{c|}{Message}                & Persentages \\ \hline
1st                      & \multicolumn{1}{c|}{no viable alternative at input `:'}  & 10\%        & \multicolumn{1}{c|}{mismatched input `.'}   & 29\%        \\ \hline
2nd                      & \multicolumn{1}{c|}{no viable alternative at input `\{'} & 7\%         & \multicolumn{1}{c|}{missing `;'}            & 25\%        \\ \hline
3rd                      & \multicolumn{1}{c|}{no viable alternative at input `)'}  & 6\%         & \multicolumn{1}{c|}{mismatched input `)'}   & 13\%        \\ \hline
4th                      & \multicolumn{1}{c|}{no viable alternative at input `('}  & 6\%         & \multicolumn{1}{c|}{mismatched input `:'}   & 4\%         \\ \hline
5th                      & \multicolumn{1}{c|}{extraneous input `\textless{}'}      & 5\%         & \multicolumn{1}{c|}{mismatched input `{]}'} & 4\%         \\ \hline
\end{tabular}
}
\end{table}

ANTLR contains an automatic recovery mechanism that allows the parser to continue after finding a syntax error.
This mechanism allows the ANTLR parser to scan the entire file as much as possible and generate a syntax tree even when encountering syntax errors. 
Relying on this mechanism, the token bag generator of MSCCD can sometimes still generate the correct token bag even when encountering syntax errors.
When using the default error recovery strategy, 
the parser will first try to perform single-token insertion and single-token deletion to recover the parser process inline. 
If it can not recover inline, the ANTLR parser will consume tokens until it finds a token that reasonably follows the current rule, which is called panic mode \cite{ANTLRBook}.
When recovered by single-token insertion or deletion, the parser will report a ``missing token" or ``extraneous input" message.
In this situation, MSCCD is able to produce the desired results.
When it turns to panic mode, the ANTLR parser will throw NoViableAltException or InputMismatchException and report a ``mismatched input" or ``no viable alternative at input" message.
If recovered successfully, MSCCD can also produce the desired results.
If the recovery fails, the token bag cannot be extracted after the error position. 
Generally, automatic recovery for ``mismatched input" is more likely to be successful than for ``no viable alternative at input".
This is because the InputMismatchException often involves more specific and localized issues,
while the NoViableAltException may involve broader context and starutural problems.
In summary, successful inline recovery does not affect token bag generation; 
however, entering panic mode may lead to the loss of token bags due to the incompleteness of the parse tree, 
and NoViableAltException is more likely to cause an impact.

For JavaScript and C++, which generated a relatively higher number of syntactic analysis errors, 
we collected all their error messages and extracted the proportions of various message types (listed in Table \ref{Table:ProportionOfErrorMessages}). 
Firstly, in terms of the proportion of errors successfully recovered inline, 
C++'s 32.14\% is higher than JavaScript's 18.05\%. 
For the two exceptions that enter panic mode, C++ is mainly constituted by InputMismatchException, accounting for 59.11\%, 
whereas JavaScript is predominantly due to NoViableAltException, accounting for 64.46\%. 
This explains why JavaScript is more sensitive to syntax errors compared to C++, 
resulting in a greater loss of token bags.


We also collected the five most frequent error messages (listed in Table \ref{Table:TopParserErrorMessages}).
As for the top frequent error messages, 
the second most common error message in JavaScript was ``no viable alternative at input `\{' ",
which usually indicates an issue with brace matching. 
Braces in JavaScript are often used to define code blocks, 
such as function bodies. 
These elements appear more frequently at the top of the syntax tree, 
and problems occurring at higher positions typically mean more information loss. 
Conversely, the second most common error in C++, ``missing `;' ", 
usually occurs at the end of an individual statement, is easy to handle, and does not lead to significant information loss.

\subsubsection{ Challenges in ANTLR Grammar Definitions}

Grammars-v4 has grammar definition files for over 100 languages that can be used directly in MSCCD. 
However, these grammar description files could benefit from further refinement to improve their reliability.
In experiments in Section \ref{Sec6.1.1}, 23 out of 48 languages showed at least errors in syntactic analysis, even in ANTLR parser construction. 
Some of these issues can be attributed to problems in the ANTLR grammar definitions. 
For instance, parser generation failures or crashes during runtime can occur due to grammar definition files not adhering to the required rules. 
In addition, issues with the grammar definition file may also lead to incorrect results in syntax analysis, such as failing to scan the entire file (whether or not errors are reported).
For users without expert-level knowledge, fixing these errors presents a significant challenge. 
Similarly, adding support for a new language requires creating ANTLR grammar definition files from scratch. 
Even if language definition files for the language already exist, 
porting them to ANTLR is not straightforward if those files are written in other grammars, such as LR grammar (left-to-right, rightmost derivation grammar),  or formats such as ENF (Backus-Naur Form) and EBNF (Extended Backus-Naur Form).
Therefore, the ability to extend MSCCD's grammar highly depends on the activity and contributions within the grammars-v4 community.


\subsection{CCFinder Series}

The CCFinder series is also known for its multilingualism, as CCfinder and CCFinderX support 5--6 languages \cite{ccfinder,CCFinderX}. 
Adding new support for them would require replacing the lexical or island-parser-like parser of the corresponding language.
Therefore, they do not fit the definition of this study on a multilingual clone detector. 
Their successor, CCFinderSW, provides a multilingual processing module that converts specific parts of the grammar definition of the targeted language into regular expressions to build a source code processor for the target language automatically. 
In previous experiments, techniques of CCFinderSW were successfully applied to 43 of the 48 tested languages \cite{CCFinderSW}. 
In the current version, CCFinderSW has built-in support for 15 languages.
Adding new language support to CCFinderSW is relatively easy, as a syntax description file need not be written. 
However, some languages whose grammar productions cannot be converted to regular expressions, such as Lua, do exist, and
CCFinderSW cannot support such languages \cite{CCFinderSW}.

\subsection{NiCad}

The newest version of NiCad supports eight languages. 
The source code processing of NiCad is based on TXL, a programming language specifically designed for software analysis and source transformation \cite{NICAD, cordy2006txl}. 
NiCad allows users to specify the analysis method for each language by TXL.
However, similar to writing an ANTLR grammar for a language from the beginning, it is not easy to describe the analysis method using TXL for developers with limited knowledge. 
Besides, the current version of TXL supports 22 languages. 
Users can reuse TXL resources of these 22 languages to specify the code analysis method of NiCad.

\subsection{Comparasion}
\begin{figure}
    \centering
    \graphicspath{{./img/}}
    \includegraphics[width=\textwidth]{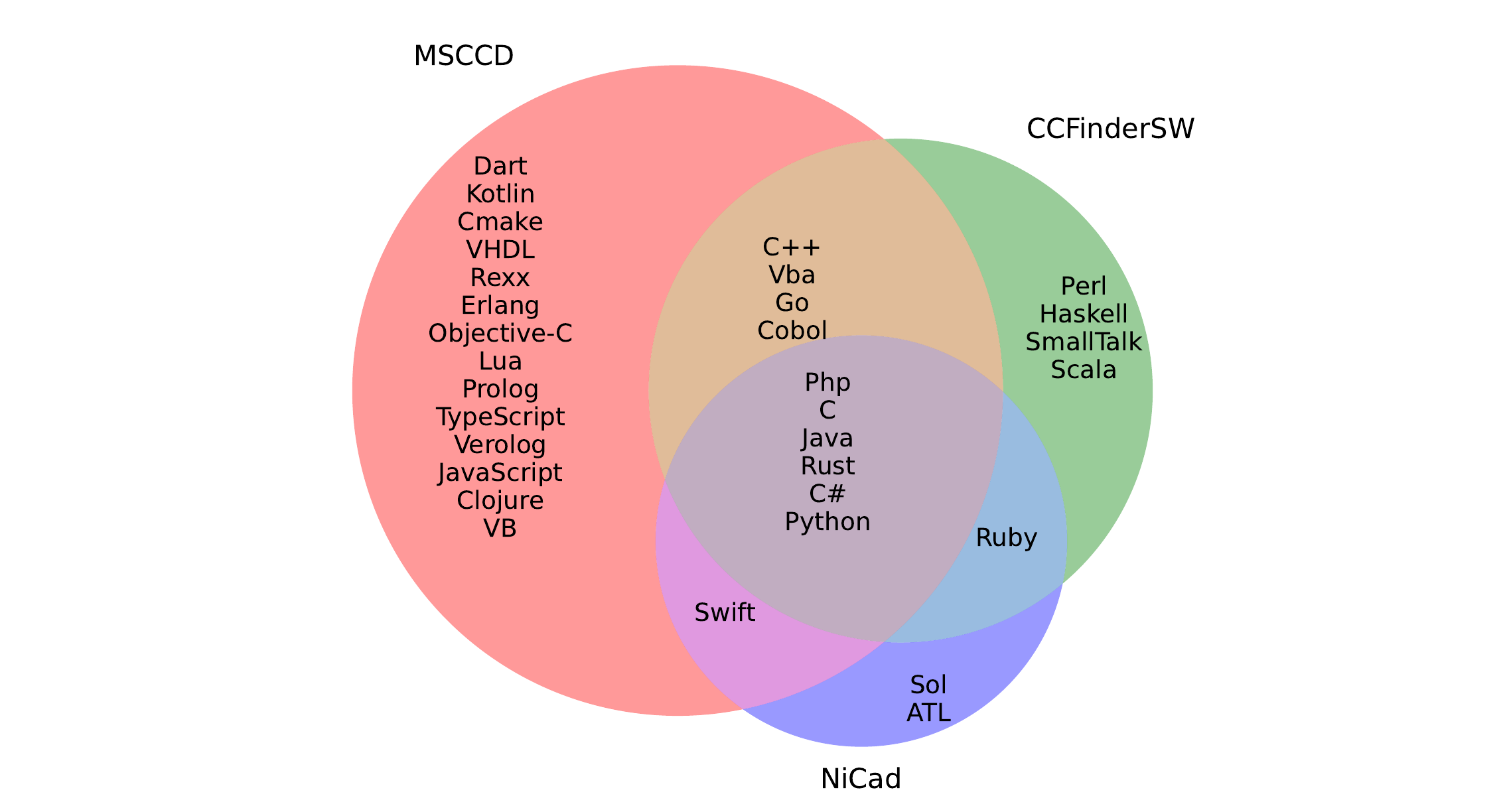}
    \caption{Currently Supported Languages of Three Multilingual Code Clone Detectors}
    \label{fig:languages}
\end{figure}

The current version of MSCCD supports the highest number of languages. 
Fig. \ref{fig:languages} shows the relationship of the current built-in language of the three multilingual code clone detectors, which allows the user to perform detection without any new configuration (including the configuration of syntax definitions). 
MSCCD supports the most significant number of languages, including 15 that were not included in the other two tools. 
With the resources of Grammars-v4, the addition of new language support to MSCCD and CCFinderSW is easy. 
Although MSCCD suffers from parser performance problems, 
it supports a much larger set of languages (25) than CCFinderSW (15) or NiCad (10).
NiCad has the lowest built-in language support and resources available (22 languages the TXL community provides). 
In addition, NiCad encounters a large number of compiler errors during the experiments, which has a significant impact on clone detection.

\noindent\fbox{
  \parbox{0.95\textwidth}{
The answer to RQ3:
MSCCD supports the maximum number of programming languages while extending CCFinderSW without existing syntax resources is more accessible. Furthermore, both MSCCD and NiCad are prone to syntax analysis failures.
  }
}

\section{Discussions}
\label{Sec7}

\subsection{What Is a Better Multilingual Code Clone Detection Approach?}

As discussed in Section \ref{Sec6}, the three main multilingual clone detectors had different language extensibility. 
Adding new language extensions to CCFinderSW was the easiest, but its approach could not be applied to some languages, such as Lua. 
Adding new language extensions to MSCCD by writing ANTLR syntax description files for a language was more complicated, 
but resource availability was the highest in this case. 
Adding new language extensions to NiCad required only writing partial grammar descriptions.
However, it was still not easy to use, and its available resources were the lowest.
The clone detection performances of the detectors were also different. 
Regarding the types of clones that can be detected, CCFinderSW supported up to T2, while MSCCD and NiCad could detect T3 clones. 
NiCad had a higher recall rate for near-miss clones than MSCCD, but its precision was less stable than MSCCD. 
Hence, no absolute leader among the three multilingual clone detectors was identified, but MSCCD was the most balanced.

Users are still required to weigh the pros and cons of using the best tool for their task requirements. 
In addition, there are more advanced or specialty, more prominent detection techniques, such as srcClone and NLC, that are yet to be extended to other languages.
Tools that use generalized tools for source code processing (MSCCD, NiCad) also suffer from parsing errors. 
To address such situations, we require the conception of a multilingual code clone detection infrastructure that can provide the easiest possible excuse and format to advance the use of official parsers, 
while providing a unified interface for clone detection without considering source code transformation.
Developers can implement their clone detection methods on that infrastructure to support multiple languages.

\subsection{Real-World Applicability of MSCCD}

This section mainly discusses the practical applications of MSCCD in the real world. 
Apart from general single-language clone detection, we attempt to outline the following application scenarios.

Scenario 1: Support for Multilingual Software Systems.
As software systems increasingly utilize multiple languages, it is preferable to use a clone detector that supports all target languages simultaneously. 
Different code clone detection tools vary in precision, granularity, result sorting strategies, and output formats. 
Standardizing these elements would incur significant costs. 
Therefore, it is advisable to avoid mixing different clone detectors in the same task. 
MSCCD can be used to address this situation.

Scenario 2: Support for Many Higher-level Applications.
Code clone detection is not the endpoint of a task but rather the starting point. 
Many tasks can be accomplished based on the results of code clone detection, such as vulnerable code detection \cite{vuddy}, automatic software repair \cite{ShapingPRSpaceWithSimilarCode}, code search \cite{kim2018facoy}, and software refactoring support \cite{higo2004refactoring}. 
When developing applications for these uses, MSCCD can be an instantly usable clone detector for the target language, playing a vital role in prototype validation and formal development. 
This allows developers to focus on developing higher-level functionalities without paying particular attention to implementing code clone detection technology, especially support for target languages.

MSCCD's most significant advantage lies in its ability to extend Type-3 level clone detection capabilities to many programming languages with minimal cost. 
The application scenarios listed above are based on this strength. 
Correspondingly, some of MSCCD's weaknesses also present challenges. 
Different tasks have varying requirements for the precision and expansiveness of clone detection tools. 
MSCCD cannot detect semantic clones and a part of Type-3 clones with very low similarity. 
It is not at the forefront regarding scalability. 
These weaknesses may render MSCCD unsuitable for specific tasks. 
Furthermore, most tasks require further development of higher-level programs based on MSCCD.
The current version of MSCCD do not provide enough APIs to support the development of these programs, requiring developers to invest effort in it.




\subsection{Threats to Validity}

\subsubsection{Comparing with Deep Learning-based Detectors}

Clone detectors based on Deep Learning (DL) can accurately detect syntactic and semantic code clones. 
However, we did not include DL-based detectors in the scope of this research's experiments because they lack generalizability, limiting their real-world usage.
In the work of Choi et al., \cite{InvestigatingGeneralizabilityOfDL}, they evaluated three DL-based detectors, including CCLearner, ASTNN, and CodeBERT\cite{li2017cclearner, ASTNN, CodeBERT}. They compared the accuracy of each detector when using the same and different benchmarks in training and testing. During their experiment, the \textit{Matthews Correlation Coefficient} (\textit{MCC}) score of all three detectors significantly decreased when using the different benchmarks for training and testing. E.g., CodeBERT performed a 0.998 \textit{MCC} score when using CodeNet for training and testing. When using CodeNet for training and BigClongBench for testing, the \textit{MCC} score decreased to 0.00432. For the syntactic code clones, their accuracy also decreases significantly. E.g., CodeBERT achieved a 0.881 \textit{MCC} score when using BigCloneBench for training and testing. However, when using the CodeNet or Google Code Jam benchmark for training, the \textit{MCC} scores for ST3 clones were 0.0039 and 0.0000778. Because the benchmarks in our work (BigCloneBench and CodeNet) are also used in their work, evaluating DL-based tools in our experiments can not gain a different result in both detection accuracy and generalizability.


\subsubsection{Configurations of the Detectors to Be Compared}

In Section \ref{Sec4} and Section \ref{Sec5}, we compared MSCCD with many state-of-the-art syntactic code clone detectors and used their default configurations as much as possible. However, according to the research of Ragkhitwetsagul et al., \cite{ragkhitwetsagul2018comparison}, the default configurations may not be the best configurations of each clone detector. Not using the optimal configuration of each tool may reduce the effectiveness of the comparison. However, we still use the default configuration of each detector as much as possible for the following reasons. First, there is a lack of research on the optimal configuration of each detector, and it is hard to research the optimal configuration for all target tools in this work. Second, most existing studies use the default configuration for evaluation. Using the default configuration makes comparing the results with existing studies possible. In addition, the optimal configuration of each tool may contain different code lengths (such as the number of tokens or lines of code). Different code length configurations will cause each tool to detect different parts of the benchmark, resulting in unfair comparisons. In the experiments in Section \ref{Sec5}, we configured the code length of each tool for the same size to ensure fairness.

\subsubsection{Bias May Introduced When Constructing the Multilingual Precision Benchmark}

In constructing the multilingual precision benchmark, we extracted 30 pairs of OJ problems for each language, divided into three similarity intervals. 
The final precision of the target tool for a specific language within a given similarity interval is determined by its ability to differentiate the code from these 30 problem pairs correctly. 
The similarity interval represents the difficulty level.
During this process, the average distribution of samples affects the effectiveness of this evaluation. 
Suppose the samples are concentrated on the right side of the similarity interval, meaning more question pairs are distributed on the higher difficulty side of the interval. In that case, the actual difficulty of this group is higher than the designated level, and the resulting accuracy might be lower than the actual level. 
Conversely, if the samples are concentrated on the left side, 
the actual difficulty of the group is lower, and the resulting accuracy might be higher.

\begin{figure}
    \centering
    \includegraphics[width=0.99\textwidth]{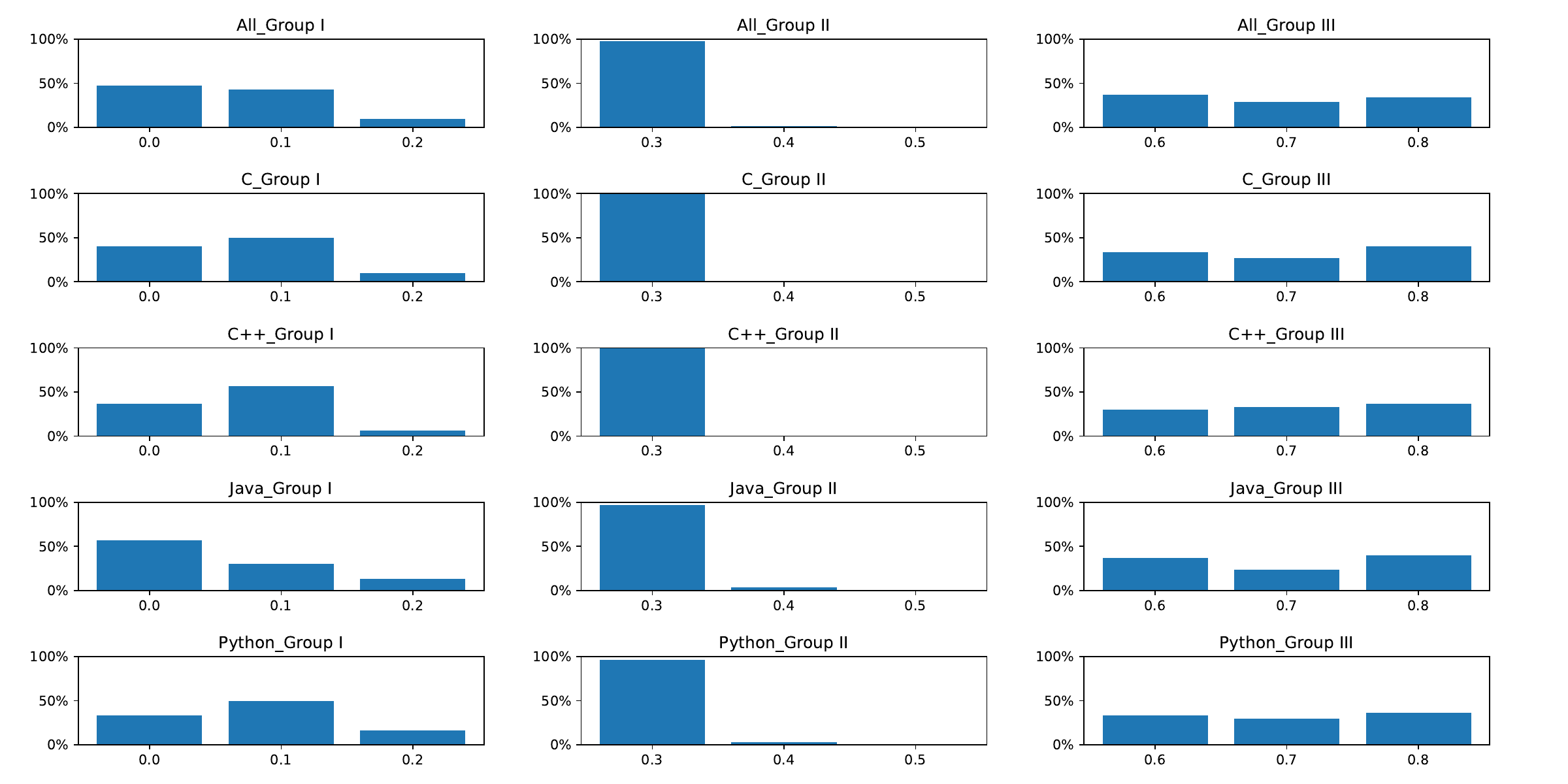}
    \footnotesize{\flushleft{ 
    \tiny
    x-axis: Problem Similarity, y-axis: Distribution in the Group
    \noindent\\
    }}
    \caption{Distribution of Similarity of OJ Problem Pairs Sampled}
    \label{fig:OJProblemSimiDistribution}
\end{figure}

We examined the distribution of the 12 sample groups in this study and created bar charts (Figure \ref{fig:OJProblemSimiDistribution}) to illustrate the distribution of samples in each group. 
The x-axis of each subplot represents the similarity of question pairs, 
and each group was divided into three intervals of 10\% for statistical purposes. 
The y-axis of each subplot shows the proportion of samples in that interval relative to the total number of samples in the group. 
Additionally, the first row of subplots represents the similarity distribution for all problem pairs. 
Firstly, the distribution characteristic of each sample group generally aligns with the overall characteristic, with no extreme cases observed. 
Groups \RomanNumeralCaps{1} and \RomanNumeralCaps{3} are relatively evenly distributed, whereas in Group \RomanNumeralCaps{2}, samples with a similarity of 30\%-40\% account for over 99\% of the total. 
This indicates that the actual difficulty of Group \RomanNumeralCaps{2} is closer to a similarity of 30\%-40\% rather than 30\%-60\%, 
suggesting that Group 2 might be a slightly more lenient group.

The similarity between the description texts of OJ problems is used to represent the similarity between the problems themselves. 
A high similarity in description texts typically indicates a high similarity between the problems. 
However, exceptions may occur in the opposite scenario. 
Since the same meaning can sometimes be conveyed using different words and sentences in natural language, 
there's a possibility that the similarity between description texts is low while the actual similarity between the problems is high. 
If this occurs, a problem pair of higher actual difficulty might be placed in a group of lower difficulty, resulting in a higher difficulty level and a lower precision. 
Nevertheless, as each group comprises 30 problem pairs, the deviation of a single problem pair is unlikely to bias the overall group significantly. 
In our evaluation experiment, both Group \RomanNumeralCaps{1} and Group \RomanNumeralCaps{2} showed high overall precision. 
Therefore, it's highly probable that many higher-difficulty problems did not infiltrate the lower-difficulty groups.

We do not advocate using the benchmark created in this study as the sole version. 
We encourage users to employ the same methodology to generate benchmarks for comparative validation using any database based on the OJ system. 
At this juncture, users should consider the abovementioned factors to avoid bias in certain groups. 
Users need to pay attention to the sample distribution of the problem sampling and consider checking the actual similarity of the problem pairs or employing alternative methods for problem similarity calculation.

\subsubsection{Contentious in the Treatment of Clones from Different Problems}

The problem groups drawn based on the similarity of the descriptive text posed challenges, 
especially for Group \RomanNumeralCaps{3}, which had the highest text similarity. 
For some problem pairs, whether the codes from different problems should not be treated as a clone is controversial. 
Table \ref{Table:SimilarProblems} shows descriptions and possible solutions of two similar problems from CodeNet. 
Both of these problems involve operations on two given line segments, 
where problem \#02295 is about finding the intersection coordinates of the two line segments, and problem \#02296 deals with finding the shortest distance between them. 
A decomposition of the solutions to these problems reveals that they both include the sub-goal of "determining whether two line segments intersect." 
This means that although the overall objectives of the two problems are different, they share a common sub-objective. 
This similarity is also reflected in the submitted code: the code segments accomplishing the task of "determining whether two line segments intersect" are likely to have a high degree of similarity. 
For clone detection tools like MSCCD and SourcererCC, which have predetermined detection granularity and detect code clones between well-defined code blocks (or entire files), The actual comparison in this experiment is between the similarity of files, which are unaffected. 
However, tools like CCFinderSW, which do not have predetermined detection granularity and detect clones by attempting to match similar subsequences, are prone to false positives. 
In other words, the detector reports highly similar subsequences rather than similar file pairs. 
Some of these results are marked as clones by the benchmark's clone matcher because the file coverage exceeds 70\%. 
From the perspective of these types of tools, their clone reports should not necessarily be considered false positives.

A potential solution is to construct a more fine-grained benchmark. 
In this study, our benchmark is file-level, 
encompassing a large number of program codes composed of multiple functions. 
If we were to use files written with a single function to build a function-level benchmark, 
it could circumvent the conflict issue described above. 
However, we are still determining whether sufficient single-function files exist in the original data to construct such a benchmark. 
At the same time, using only single-function files might introduce new biases due to changes in average length and other factors, which is also a matter of concern.

\begin{table}[]
\caption{Description and Possible Solution of Two Similar Problems from CodeNet}
\label{Table:SimilarProblems}
\fontsize{8pt}{12pt}\selectfont
\resizebox{\textwidth}{!}{



\begin{tabular}{l|l|l}
\hline
\multirow{3}{*}{\begin{tabular}[c]{@{}l@{}}Problem \\ \#2295\end{tabular}} & Title             & Cross Point                                                                                                                                                                                                                                               \\ \cline{2-3} 
                                                                           & Description       & \begin{tabular}[c]{@{}l@{}}For given two segments s1 and s2,\\ print the coordinate of the cross point of them.\\ s1 is formed by end points p0 and p1, \\ and s2 is formed by end points p2 and p3.\end{tabular}                                         \\ \cline{2-3} 
                                                                           & Possible Solution & \begin{tabular}[c]{@{}l@{}}\textbf{1. Check if the segments intersect.}\\ 2. Output the coordinates of the cross\end{tabular}                                                                                                                                      \\ \hline
\multirow{3}{*}{\begin{tabular}[c]{@{}l@{}}Problem \\ \#2296\end{tabular}} & Title             & Distance                                                                                                                                                                                                                                                  \\ \cline{2-3} 
                                                                           & Description       & \begin{tabular}[c]{@{}l@{}}For given two segments s1 and s2,\\ print the distance between them.\\ s1 is formed by end points p0 and p1, \\ and s2 is formed by end points p2 and p3.\end{tabular}                                                         \\ \cline{2-3} 
                                                                           & Possible Solution & \begin{tabular}[c]{@{}l@{}}\textbf{1. Check if the segments intersect.}\\ 2. Calculate the distance between a point from the one line segment\\ to the other line segment if it does not intersect.\\ 3. Use the above method to find the minimum dis\end{tabular} \\ \hline
\end{tabular}
}
\end{table}

\subsubsection{The Other Points for the Benchmarks}
The key idea of multilingual code clone detection evaluation is that 
all code pairs between accepted submissions from the same OJ problem can be considered code clones.
Doing so means we do not need to label all the clone pairs manually.
However, only code clones between files were tagged, while many programs from CodeNet consisted of multiple functions. 
This made it hard for our benchmark to evaluate block-level-only clone detectors
because they do not report any file-level code clones.
As mentioned above, constructing a function-level benchmark is helpful but challenging.

For the precision benchmark, 
since different languages may use different question sets even within the same group,
the validity of the comparison between languages was reduced. 
We are considering reducing the benchmark size to use the same set of problems across different languages.



%

\section{Related Work}
\label{Sec8}

According to Roy's survey, 
a code clone detection approach usually has eight phases, 
among which the first three steps of preprocessing, transformation, and match detection are essential \cite{roy2007survey}. 

The target code is partitioned and transformed while preprocessing and transforming to another intermediate representation that can calculate similarity. 
These phases require syntax information of the target language to remove unnecessary parts (comments and embedded code) and transformations. 
Therefore, most clone detectors only support a few languages \cite{MSCCD}. 
To solve this problem, some researchers have proposed language processing mechanisms that allow clone detectors to support new languages more easily \cite{NICAD,CCFinderSW,MSCCD}. 
Roy and Cordy proposed a flexible pretty-printing and code normalization approach based on TXL, a special-purpose programming language \cite{cordy2006txl}, and implemented the NiCad clone detector \cite{NICAD}.
Users are allowed to add new language support for NiCad by providing a source code analysis method written by TXL.
Although writing TXL methods is more accessible than writing parser generator grammar, 
it is still difficult for users with limited knowledge. 
To date, NiCad does not support a number of languages.
CCFinderSW \cite{CCFinderSW} implements a multilingual lexer to generate a token sequence for clone detection \cite{CCFinderX}. 
The main idea is to convert the grammar rules of comment, identifier, keyword, and literal into regular expressions.
Since there are only four syntax rules to focus on, 
adding language support to CCFinderSW is the easiest. 
However, some syntax rules cannot be converted to regular languages, such as the comment of Lua; thus, it is not supported by CCFinderSW. 
In addition, the approach of CCFinderSW can only generate token sequences.
Currently, it only supports Type-2 clone detection, which is extremely difficult to extend in the future. 
The MSCCD source code processing approach can generate token bags and can be simply extended to token sequences and PTs. 
More accurate clone detection techniques are expected to be implemented using MSCCD.

For match detection, the clone detector compares partitioned code to find similar code segments. 
These detection approaches can be categorized by the level of intermediate representation as 
text-based \cite{NICAD,nasehi2007source}, token-based \cite{SourcererCC,MSCCD,CCAligner,ccfinder,basit2007efficient,siamese}, tree-based \cite{jiang2007deckard,baxter1998cloneast,koschke2006clone}, metrics-based \cite{saini2018oreo}, and graph-based \cite{ccsharp,kamalpriya2017enhancing}. 
NiCad\cite{NICAD}, iClones\cite{iClones}, Deckard\cite{jiang2007deckard}, SourcererCC\cite{SourcererCC}, Oreo\cite{saini2018oreo}, and CloneWorks\cite{svajlenko2017fast} are some of the well-known Type-3 level clone detectors.
Some detectors focus on individual metrics, 
such as NIL\cite{NIL} and CCAligner\cite{CCAligner} that focus on detecting large-variance and large-gap clones, 
while SAGA\cite{SAGA} extends the scalability of clone detection to 550MLOC.
Although there is a gap between MSCCD and the latest detectors in terms of near-miss clone detection and scalability, 
it supports the largest number of programming languages and also has above average detection accuracy and scalability. This makes MSCCD the most balanced clone detector. 
We expect more clone detection techniques to have multi-language implementations in the future.

All the aforementioned tools are designed mainly for syntactic code clones, and there is another class of techniques designed to detect semantic code.
SrcClone analyzes semantic similarity by program slicing technique \cite{srcClone}.
Some AI-based clone detectors are also frequently used to detect semantic code clones \cite{wu2020scdetector,YU_IST_SEMANTICCLONE,FangISSTA2020SemanticClone}.
SCDetector performs well by combining the information about the token and the control flow graphs using a neural network \cite{wu2020scdetector}.
Yu et al. proposed an approach to learn semantics based on  CFG (Control Flow Graph) or PDG (Program Dependency Graph), while Fang et al. proposed to learn semantic by applying fusion embedding techniques.
However, according to research from Choi et al. and Liu er al. \cite{InvestigatingGeneralizabilityOfDL,CanNeuralCloneDetectionGeneralizeToUnseenFunctionalitiesƒ}, AI-based clone detection tools have limitations in their generalization capabilities. 
Therefore, we didn't include these AI-based approaches in our experiments. 
The existing multilingual code clone detectors (MSCCD, NiCad, CCFinderSW) cannot detect semantic clones.
Besides, neither code slicing nor control flow graph generation has a multilinguistic approach.
Therefore, multilingual semantic clone detection is not supported at the moment. 


Unlike the above-introduced approaches, cross-language code clone detectors aim to find code clones between different programming languages.
Perez et al. proposed an approach to detect code clones between Java and Python that learn token-level vector representations by an LSTM-based neural network \cite{perez2019cross}.
LICCA can detect clones between five languages by generating a high-level representation of code from the SSQSA platform \cite{LICCA_CLCCD}.
CLCDSA can detect cross-languages without creating other intermediate representations \cite{CLCDSA_CLCCD}.
It is implemented by learning and comparing the similarity of features.
Currently, there are no multilingual cross-language clone detectors. 
However, the source code processing approach of MSCCD can support multiple intermediate representations of code, such as PT.
Once a PT-based cross-language clone detection approach is proposed, it can be easily added to a multilingual implementation.

Traditional and cross-language code clones assume that the targeted software is developed in a single language.
However, a large amount of software is developed in multiple languages for compatibility with older systems or the need to take advantage of different features of different languages simultaneously. 
These include many source code files written in multiple languages (the most typical ones are HTML-JavaScript and Java server pages) \cite{Rajapakse2007, Tariq2013}. 
A clone pair can be defined such that both sides are composed of multiple languages as a language-mixed code clone. 
Nakamura et al. proposed an approach to detect language-mixed code clones between web systems \cite{Nakamura2016}. 
Besides, some frequently used language groups are supported, such as Python and Ruby.
One of our future challenges is to extend MSCCD for language-mixed code clone detection using a technique such as the island grammars.

An enduring problem for benchmarking code clone detectors for recall and precision exists.
Since the development of such a benchmark requires marking all correct code clones in all code pairs, there are no automated methods of marking code clones (perfect clone detector), 
and the workload for manually checking code clones is far beyond the limit.
Researchers initially evaluated code clone detectors by comparing their targeting at large-scale open-source software, such as Linux and NetBSD \cite{jiang2007deckard,ccfinder,cp-miner}.
Bellon et al. created the first large-scale benchmark to visually judge whether a code clone was detected \cite{bellon2007comparison}.
The most frequently used benchmark is BigCloneBench \cite{BCB,BCE} created by Scajlenko and Roy.
BigCloneBench contains over 7,800,000 pairs of Java clones and can report recalls in each of the six categories: T1, T2, VST3, ST3, MT3, and WT3/T4.
Svajlenko and Roy also published a mutation and injection framework to evaluate the recall of clone detectors for C \cite{Svajlenko2021}.
Some studies have attempted to advance the evaluation of precision \cite{Precision2019,saini2019towards}.
However, current precision evaluations still require random sampling and manual judgment.
Al-Omari et al. proposed SemanticCloneBench \cite{semanticclonebench}, which collects semantic code clones from Stack Overflow answers.
SemanticCloneBench supports Java, C, C\#, and Python. Alam et al. proposed GPTCloneBench, which is based on SemanticCloneBench and GPT-3 model \cite{GPTCloneBench}. GPTCloneBench provides more functional code examples in higher quality. However, these benchmarks were not designed for evaluating syntactic code clone detectors.

In Section \ref{Sec5} of this study, we created a benchmark to evaluate recall and precision automatically for four languages by using the project CodeNet \footnote{\url{https://github.com/IBM/Project_CodeNet}}, which included more than 13,000,000 submissions from AIZU Online Judge and AtCoder.
Competitive programming codes have always been used for evaluating deep-learning-based code clone approaches \cite{ASTNN,DeepSim,AST_CLCCD}; however,
our study is the first to use such an approach for syntactic clone detection evaluation.

\section{Conclusions and Future Work}
\label{Sec9}

This work addressed the need for existing multilingual code clone detectors. 
We proposed a code block extraction method based on the ANTLR parser generator and implemented it into MSCCD.
MSCCD can support the most popular programming languages and detect T3 clones.
Furthermore, we constructed the first multilingual syntactic code clone detection evaluation benchmark, which measures precision and recall for C, C++, Java, and Python.
According to the results of evaluation experiments, MSCCD supports the most languages among the existing tools and is at the same level of detection performance as the existing tools. 
Although MSCCD is already the most balanced multilingual clone detector, it still needs help with problems like parser errors. 
An intermediate layer for multilingual clone detection can be implemented in the future to promote the use of official parsers and to enable various detection techniques to be implemented directly in more languages.

In the future, we plan to continue improving the performance of the evaluation benchmark for multilingual clone detection and explore the best implementation of the middle layer.

\section*{Acknowledgement}
This work was supported by JST, PRESTO Grant Number JPMJPR21PA, Japan. Also, this work was supported by JSPS KAKENHI Grant Numbers JP20K11745, JP22K11975, JP23K11046 and JP24K02923.
\bibliographystyle{elsarticle-num} 
\bibliography{MSCCD_JSS}

\textbf{Wenqing Zhu} is a Ph.D. student at the Graduate School of Informatics, Nagoya University. He received his Master's degree from Nagoya University in 2022 and his B.E. from Northeast Normal University. He is interested in program analysis.

\textbf{Norihiro Yoshida} received his B.E. in Artificial Intelligence from the Kyushu Institute of Technology in 2004 and his Master of Information Science and Technology and Ph.D. in Information Science and Technology from Osaka University under the supervision of Prof. Katsuro Inoue in 2006 and 2009, respectively. He is a professor at Ritsumeikan University. He was an associate professor in the Graduate School of Informatics at Nagoya University from 2014 to 2022 and an assistant professor in the Graduate School of Information Science at the Nara Institute of Science and Technology (NAIST) from 2010 to March 2014. His research interests include program analysis and software development environments. He is a member of the IEEE, the IEEE Computer Society, and the ACM.

\textbf{Toshihiro Kamiya} received his M.Eng. in 1999 and Ph.D.Eng. in 2001 from the Department of Informatics and Mathematical Science at Osaka University. From 2001 to 2004, he served as a Research Associate in the Intelligent Cooperation and Control Group at RPESTO, a part of the Japan Science and Technology Corporation. He then worked as a Researcher at the National Institute of Advanced Industrial Science and Technology from 2005 to 2009. He held the position of Associate Professor at Future University Hakodate from 2010 to 2014. Currently, he is a Professor at Shimane University. His research interests include program comprehension, as well as static and dynamic program analysis. He is a member of IPSJ and IEEE.

\textbf{Eunjong Choi} is an assistant professor at Faculty of Information and Human Sciences in Kyoto Institute of Technology, Japan since March 2019. Before joining Kyoto Institute of Technology, She was an assistant professor at Nara Institute of Science and Technology from April 2016 to March 2019 and at Osaka School of International, Osaka University from April 2015 to March 2016. Her research interests are in the areas of software engineering, in particular, reused code management/detection and refactoring support. 

\textbf{Hiroaki Takada} is a professor at the Institutes of Innovation for Future Society, Nagoya University. He is also a professor and the Executive Director of the Center for Embedded Computing Systems (NCES), the Graduate School of Informatics, Nagoya University. He received his Ph.D. in Information Science from the University of Tokyo in 1996. He was a Research Associate at the University of Tokyo from 1989 to 1997 and was a Lecturer and then an Associate Professor at Toyohashi University of Technology from 1997 to 2003. His research interests include real-time operating systems, real-time scheduling theory, and automotive embedded systems. He is a fellow of IPSJ and JSSST and is a member of IEEE, IEICE, and JSAE.

\end{document}